\documentclass[aps, twocolumn ,  prb, showpacs]{revtex4}
\usepackage{amsmath,amssymb}
\usepackage{subfigure}
\usepackage{epsfig}
\usepackage{graphicx}
\begin{document}
\author{Y. Hamdouni\footnote{Electronic address: hamdouni@ukzn.ac.za }}
 \affiliation{School of Physics,
University of KwaZulu-Natal, Westville Campus, Durban 4001 , South
Africa}
\author{F. Petruccione}
\affiliation{School of Physics, University of KwaZulu-Natal,
Westville Campus,  Durban 4001 , South Africa}

\title{Time evolution and decoherence of a spin-$\frac{1}{2}$ particle coupled to a
 spin bath in thermal equilibrium}
\pacs{ 03.65.Yz, 03.67.Lx, 73.21.La, 75.10.Jm}

\begin{abstract}
 The time evolution of a spin-$\frac{1}{2}$ particle under
the influence of a locally applied external magnetic field, and
interacting with anisotropic spin  environment in thermal
equilibrium at temperature $T$ is studied. The exact analytical form
of the reduced density matrix of the central spin is calculated
explicitly for finite number of bath spins. The case of an infinite
number of environmental spins is investigated using the convergence
of the rescaled bath operators to normal Gaussian random variables.
  In this limit, we  derive the analytical form of the components of the Bloch vector  for
   antiferromagnetic interactions within the bath, and we investigate
the short-time and long-time behavior of reduced dynamics. The
 effect of the external magnetic field, the anisotropy and the temperature of the bath on the decoherence of the central spin is discussed.
\end{abstract}

\maketitle \section{Introduction}

The loss of quantum coherence due to unavoidable interactions of quantum systems with the
surrounding environment  is known as  decoherence. It represents the main obstacle to quantum
computing and quantum information processing~\cite{1,2,3}.
 The environment
destroys quantum interferences of the central system within time scales much
shorter than those typically characterizing dissipation~\cite{4}.
  The unwanted effect of decoherence reduces the advantages of quantum computing
 methods by producing errors in their outcomes. Different strategies,  such as
 error-correcting codes, are adopted to overcome this difficulty~\cite{5,6,7,8}.
 Great scientific effort has been devoted to
 the understanding of the process of
decoherence in  quantum systems, mainly  focused  on
solid state spin nanostructures. These systems seem
to be the most promising candidates  that can
be efficiently used in quantum information processing and computation~\cite{9,10,11}.

Several models were proposed to study decoherence of single and multi-spin systems interacting with
a surrounding environment~\cite{12}.
  Very often, the derivation of the reduced dynamics  involves
complications and difficulties that can  be
 overcome in many cases by  making recourse to approximation techniques. In particular, the Markovian
  approximation together with the  master equation approach
turns out to be very useful~\cite{13,14}. However, any approximation method is inevitably based on some assumptions which do not necessarily
reflect the actual properties of the composite system.
Moreover, many realistic spin systems exhibit non-Markovian behavior for which the standard derivation of
the master equation  ceases to be applicable. The non-Markovian dynamics of a central spin-system coupled to a spin environment
has been investigated by many authors~\cite{15,16,17,18,19}.

 In general, the course of the decoherence process depends on
the intrinsic properties of the bath such as temperature, polarizations, and quantum
fluctuations. At low  environmental temperatures, the dominant effect arises from
the contributions of  localized modes such as nuclear spins~\cite{20}.
In quantum dots, the  decoherence of the  central spins is mainly caused by
the hyperfine coupling with the surrounding nuclear spins. The effect of bath polarizations and external
magnetic fields on the decoherence of electron spins in quantum dots  has been investigated by Zhang {\it et al}~ \cite{21}.

In the following paper we study the dynamics of a spin-$\frac{1}{2}$
particle interacting with a large spin environment  in thermal
equilibrium. In Sec.~II we introduce the model Hamiltonian together
with the initial states of the central spin and the environment.
In Sec.~III, we calculate the exact time evolution operator of the
composite system  and we derive the reduced density matrix of the
central spin. Sec.~IV is devoted to the case of an infinite number of
spins in the bath.  We study the long-time behavior as well as the short-time behavior of the reduced
density matrix, and we discuss the effect of the magnetic field and the bath temperature on decoherence. A short conclusion ends the paper.
\section{The model}

We consider a central spin-$\frac{1}{2}$ particle  coupled to a
 spin  bath  composed of $N$ interacting spin-$\frac{1}{2}$
  particles  in thermal equilibrium at temperature
 $T$.
  The spin operators corresponding to the central system are
 denoted by $S^0_i$ with $i=x, y, z$, those associated with the bath
 constituents are denoted by $ S^k_i$, where $k=1,2,...,N$ and
 $i=x,y,z$.  We
 assume that the central system as well as  every spin in the bath couples to all other spins through
 long-range  anisotropic Heisenberg  interactions. Moreover, an external magnetic field  of controlled strength $\mu$
 is locally applied to the central spin along the
 $z$ direction. Under the above assumptions, the model
 Hamiltonian can be written as
 \begin{equation}
 H=H_S+H_{SB}+H_{B}
 \end{equation}
 where $H_S$ and $H_{B}$ are, respectively, the Hamiltonian operators
 of the central spin and the  surrounding
 environment. The coupling between the open system and the bath is described by the Hamiltonian $H_{SB}$. Explicitly, we have
 \begin{align}
 H_S&= 2 \mu S^0_z, \label{h1}\\
 H_{SB}&=\frac{2\gamma}{\sqrt{N}} \
 S^0_z \sum\limits^{N}_{i=1}{ S^i_z}+
 \frac{2\alpha}{\sqrt{N}}\Bigl[S^0_x\sum\limits^{N}_{i=1}{S^i_x}+S^0_y\sum\limits^{N}_{i=1}{S^i_y}\Bigl], \label{h2}\\
 H_B&=\frac{g}{N}\Bigl[\sum\limits^{N}_{i\ne j}{\Bigl(S^i_x S^j_x+S^i_y
 S^j_y\Bigl)}+\Delta \sum\limits^{N}_{i\ne j}{S^i_z
 S^j_z}\Bigl],\label{h3}
\end{align}
where $ \gamma$ and  $\alpha$ are the coupling constants of the
central spin to the environment, $g$ stands for the strength of
interactions of  spins in the bath,  and $\Delta$ is the anisotropy
constant.  The coefficient 2 in front of $\mu$, $\gamma$ and
$\alpha$ in Eqs.~(\ref{h1}) and (\ref{h2})
 is introduced for later
convenience. Furthermore, we have rescaled the above interaction strengths
 with appropriate powers of the number
of spins in the environment in order to ensure good thermodynamical behavior, namely an
 extensive free energy. Obviously, a more realistic model would include
 site-dependent interactions.

  Note that in the case where $\gamma=0$, $H_{SB}$ reduces to Heisenberg $XY$
 Hamiltonian which was recently used to model the coupling of one and two qubits
  to star-like environments~\cite{22,15,16,17}. Moreover, when
  $\gamma=\alpha$ we simply have
 $H_{SB}=\frac{\alpha}{\sqrt{N}}\vec{S}^0\sum\limits_{i=1}^N\vec{S^i}$, which should be compared with the Hamiltonian of the
hyperfine contact coupling of  electron spin
 to the nuclear spins in quantum dot~. In Ref.~\onlinecite{23}, the Hamiltonian
$h_B=\sum_{i>j} g_{ij}({\vec{S}^i\vec{S}^j-3S^i_zS^j_z})$
   was used   to model the intra-bath dipolar coupling between nuclear spins in quantum dot.
       If we assume uniform coupling between nuclear spins, i.e. all the $g_{ij}$ are the same,
       then the operator $h_B$ (with rescaled coupling
    constant) becomes equivalent to $H_B$ in the case where $\Delta=-2$.
 It should also be noted that the  bath Hamiltonian  $H_B$   is very
 close to that of the isotropic Lipkin-Meshkov-Glick model~\cite{24,25}.
  There, the  magnetic field globally applied to all spins  plays  the role of the anisotropy present
 in our model. This can be better seen by applying  mean field approximation to the longitudinal term of $H_B$.

The Hamiltonian operators $H_{B}$ and $H_{SB}$ can be rewritten in
terms of the lowering and raising operators $S^i_{\pm}=S^i_x\pm i
S^i_y$ as follows
\begin{align}
 H_{SB}&=\frac{2\gamma}{\sqrt{N}}\ S^0_z
 \sum\limits^{N}_{i=1}{S^i_z}+
 \frac{\alpha}{\sqrt{N}}\Bigl[S^0_+\sum\limits^{N}_{i=1}{S^i_-}+S^0_-\sum\limits^{N}_{i=1}{S^i_+}\Bigl], \\
 H_B&=\frac{g}{2N}\Bigl[\sum\limits^{N}_{i\ne j}{\Bigl(S^i_+ S^j_-+S^i_-
 S^j_+\Bigl)}+2 \Delta \sum\limits^{N}_{i\ne j}{S^i_z
 S^j_z}\Bigl].
\end{align}
 By introducing the total angular momentum of the bath
 $\vec{J}=\sum\limits_{i=1}^{N}\vec{S}^i$, together with  the corresponding lowering and
 raising operators $J_\pm$, it is possible to put the above Hamiltonian
 operators into the following form
 \begin{align}
 H_{SB}&=\frac{2\gamma}{\sqrt{N}}
  S^0_z J_z+
 \frac{\alpha}{\sqrt{N}}\Bigl[S^0_+ J_- + S^0_- J_+\Bigl], \label{h4}\\
 H_B&=\frac{g}{2N}\Bigl[K +2\Delta J_z^2 -\frac{(2+\Delta)N}{2} \Bigl].\label{h5}
\end{align}
Here, $J_z$ is the $z$-component of the total angular momentum $J$, and
we have introduced the  operator $K=J_+J_-+J_-J_+$. From here on, we shall neglect the constant $(2+\Delta)g/4$
appearing in the expression of $H_B$
since it has no effect on the dynamics of the system. This can be
 done by redefining the energy origin of the spectrum of the
bath Hamiltonian.

 The spin spaces corresponding to the central spin and the
environment are  given by $\mathbb{C}^2$ and
$(\mathbb{C}^2)^{\otimes N}$, respectively. The latter space can be
decomposed as a direct sum of subspaces $\mathbb C ^{d_j}$ each of
which has a  dimension equal to $d_j=2 j+1$ where $0\le j \le
\frac{N}{2}$~\cite{16}( we take $N$ even), namely $(\mathbb
C^2)^{\otimes N}=\bigoplus \limits_{j=0}^{\frac{N}{2}}\nu(N,j)
\mathbb C^{d_j}$. The degeneracy $\nu(N,j)$ is given by~\cite{26}
\begin{equation}\nu(N,j)=\frac{2j+1}{\frac{N}{2}+j+1}
\frac{N!}{(\frac{N}{2}-j)! (\frac{N}{2}+j)!}.
\end{equation}
 It is worth noting that
 the bath Hamiltonian can be expressed in terms of the operators
$J^2$ and $J_z$ as $H_B=\frac{g}{N}[J^2+(\Delta-1)J_z^2]$.
Therefore, the operator  $H_B$ is diagonal in the standard basis of
$(\mathbb C^2)^{\otimes N}$ formed by the common eigenvectors  of
$J^2$ and $J_z$ which we denote by $|j,m\rangle$
 where $-j\le m\le j$. In this basis, the eigenvalues of the operator $K$
 are simply given by $2(j(j+1)-m^2)$ (we set $\hbar=1$).
\section{Reduced dynamics of the central spin}
In this section we derive the exact time evolution of the central
spin for finite number of environmental spins. As usual, we introduce
the time evolution operator ${\bf U}(t)=\mathrm e^{-i H t}$ together with
the total density matrix operator of the spin-bath system,
$\rho_\mathrm{tot}(t)$. The initial value of the latter
 is denoted by
$\rho_\mathrm{tot}(0)$.
The evolution in time of the composite system is unitary, its density matrix at any  moment of time
is given by
 \begin{equation}
\rho_\mathrm{tot}(t)={\bf U}(t) \rho_\mathrm{tot}(0) {\bf U^\dag} (t).\label{ev}
\end{equation}
 The reduced density matrix of the central spin can be
calculated by tracing  $\rho_\mathrm{tot}(t)$ with respect to
the environmental degrees of freedom, namely
\begin{equation}
\rho(t)= \mathrm{tr}_B\{\rho_\mathrm{tot}(t)\}\label{tr}.
\end{equation}
This can be explicitly  written in terms of bath states as
\begin{equation}
\rho(t)= \sum\limits_{j,m}\nu(N,j) \langle j,m|\rho_\mathrm{tot}(t)|j,m\rangle.
\end{equation}
 In order to solve the time evolution problem (\ref{ev}), one needs to
 calculate the exact analytical form of  $\bf{U}(t)$
and to specify the  initial density matrix.
\subsection{Initial conditions}
 Initially, the central spin is assumed to be uncorrelated with the environment. The corresponding total density matrix is  given by
the  direct product $\rho_\mathrm{tot}(0)=\rho(0)\otimes\rho_B$
where $ \rho(0)$ and $\rho_B$ are, respectively, the initial density
matrices of the central spin and the bath. In the standard basis
composed of the eigenvectors $|-\rangle $ and $|+\rangle$ of the
operator $S^0_z$, $\rho_S(0)$ takes the  general form
\begin{equation}
\rho(0)=\begin{pmatrix}\rho_{11}^0& \rho_{12}^0\\
 \rho_{12}^{0*}& \rho_{22}^0\end{pmatrix},\end{equation}
where $\rho_{11}$ and $\rho_{22}$ are positive real numbers which
satisfy $\rho_{11}+\rho_{22}=1$. For instance, if at $t=0$ the
central system was in the state
\begin{equation}
|\psi(0)\rangle=a |-\rangle + b |+\rangle,
\end{equation}
where $a$ and $b$ are complex numbers satisfying $|a|^2+|b|^2=1$,
then $\rho_{11}^0=|a|^2$ and $\rho_{12}^0= a b^*$.\\
Alternatively, $\rho(0)$ can be expressed in terms of  the components of the Bloch
vector $\vec{\lambda}=(\lambda_1,\lambda_2,\lambda_3)$  as
\begin{equation}
\rho(0)=\frac{1}{2}\begin{pmatrix}1-\lambda_3(0)& \lambda_1(0)-i\lambda_2(0)\\
\lambda_1(0)+i\lambda_2(0)&1+\lambda_3(0)\end{pmatrix},\end{equation}
with the condition $|\vec{\lambda}|\leq 1$; the equality holds for
pure initial  states. We shall use both representations of the
density matrix throughout the paper.

 At $t=0$, the spin bath is taken  in thermal equilibrium at finite temperature $T$.
 Its density matrix  is given by the
 Boltzmann distribution
\begin{equation}
\rho_B=\frac{\mathrm e^{-\beta H_B}}{Z_N},
\end{equation}
where $\beta=1/T$ (we set $k_{B}=1$), and $Z_N=\mathrm {tr}_B
\mathrm e^{-\beta H_B}$ is the partition function of the bath.
Clearly,  $ \rho_B$ is diagonal in the standard basis
$\{|j,m\rangle\}$ from which it  follows that~\cite{27}
\begin{equation}
\langle j,m|\rho_B|j, m\rangle =\frac{1}{Z_N} \mathrm e^{-
\frac{g\beta}{N}[j(j+1)+(\Delta-1)m^2]}, \end{equation} and
\begin{equation}
Z_N=\sum_{j,m}\nu(N,j)\ \mathrm e^{- \frac{g\beta}{N}[j(j+1)+(\Delta-1)m^2]}
\label{par}.
\end{equation}
In the case of the isotropic Heisenberg model, i.e. when $\Delta=1$, the
above expression simplifies to
\begin{equation}
Z_N=\sum_{j}\nu(N,j) (2j+1)\  \mathrm e^{- \frac{g\beta}{N}[j(j+1)]}.
\end{equation}
In the extreme case of an infinite temperature ($\beta\to 0$), the
 density matrix of the bath reads
\begin{equation}
\rho_B(T=\infty)=\frac{{\bf1}_B}{2^N} ,\end{equation} which corresponds to
a completely unpolarized spin bath. In the previous expression ${\bf1}_B$ stands for the unity matrix in the bath space.
\subsection{Time evolution operator}
Let $U_{ij}$ denote the components of the time evolution operator
${\bf U}$ in the basis $\{|-\rangle,|+\rangle\}$ corresponding the
central system space. Therefore, we can write
\begin{align}
{\bf U}|-\rangle=U_{11}|-\rangle+U_{21}|+\rangle, \label{u1}\\
{\bf U}|+\rangle=U_{12}|-\rangle+U_{22}|+\rangle.\label{u2}
\end{align}
On the other hand, the operator ${\bf U}$ satisfies  the
Schr\"odinger equation
\begin{equation}
i \frac{d}{dt} {\bf U}|\pm\rangle= H {\bf U}|\pm\rangle.\label{sch}
\end{equation}
Substituting  Eq.~(\ref{u1}) into Eq.~(\ref{sch}) yields the following
 system of coupled differential equations
\begin{align}
i\dot{U}_{11}&=\Bigl[-\Bigl(\mu + \frac{\gamma J_z}{\sqrt{N}}\Bigl)+\frac
{g}{2N}\Bigl(K+2 \Delta J_z^2\Bigl)\Bigl]U_{11}+\frac{\alpha
J_+}{\sqrt{N}}U_{21},
\\
i\dot{U}_{21}&=\frac{\alpha J_-}{\sqrt{N}}U_{11}+\Bigl[\Bigl(\mu +
\frac{\gamma J_z}{\sqrt{N}}\Bigl)+\frac {g}{2N}\Bigl(K+2 \Delta
J_z^2\Bigl)\Bigl]U_{21}.
\end{align}
Similarly, from Eq.~(\ref{u2}) and Eq.~(\ref{sch}) we obtain
\begin{align}
i\dot{U}_{22}&=\Bigl[\Bigl(\mu + \frac{\gamma J_z}{\sqrt{N}}\Bigl)+\frac
{g}{2N}\Bigl(K+2 \Delta J_z^2\Bigl)\Bigl]U_{22}+\frac{\alpha
J_-}{\sqrt{N}}U_{12},
\\
i\dot{U}_{12}&=\frac{\alpha J_+}{\sqrt{N}}U_{22}+\Bigl[-\Bigl(\mu +
\frac{\gamma J_z}{\sqrt{N}}\Bigl)+\frac {g}{2N}\Bigl(K+2 \Delta
J_z^2\Bigl)\Bigl]U_{12}.
\end{align}
Since  ${\bf U}(0)={\bf 1}_2 \otimes {\bf 1}_B$, one gets the
 initial conditions
\begin{equation}U_{11}(0)=U_{22}(0)={\bf 1}_B,\ \
U_{12}(0)=U_{21}(0)=0.\label{initial}
\end{equation}
 The difficulty with solving the  above set of differential equations resides in
  the fact that the coefficients of the operator variables
  $U_{21}$ and $U_{12}$ are not diagonal and do not commute with those of $U_{11}$ and $U_{22}$.
   Nevertheless, as we shall see bellow, this problem can be overcome by transforming
   these equations
   into new ones involving commuting diagonal operators. Indeed, by making use of  the
  change of variables (see Ref.~\onlinecite{19} for a similar method)
  \begin{align}
U_{11}&= \mathrm e^{-i[-(\mu + \frac{\gamma J_z}{\sqrt{N}})+\frac{g}{2N}(K+2 \Delta J_z^2)]t} \widetilde{U}_{11},\label{tr1}\\
U_{21}& = J_-\mathrm e^{-i[-(\mu + \frac{\gamma J_z}{\sqrt{N}})+\frac
{g}{2N}(K+2 \Delta J_z^2)]t} \widetilde{U}_{21},\label{tr2}\\
U_{22}&= \mathrm e^{-i[(\mu + \frac{\gamma J_z}{\sqrt{N}})+\frac{g}{2N}(K+2 \Delta J_z^2)]t} \widetilde{U}_{22},\label{tr3}\\
U_{12}& = J_+ \mathrm e^{-i[(\mu + \frac{\gamma J_z}{\sqrt{N}})+\frac
{g}{2N}(K+2 \Delta J_z^2)]t} \widetilde{U}_{12} ,\label{tr4}
\end{align}
and taking into account the  commutation relations
\begin{equation}
[J_z,J_\pm]=\pm J_\pm, \quad [J_z^2,J_\pm]=\pm
J_\pm(2J_z\pm1)\nonumber
\end{equation}
and
\begin{equation}
[K,J_\pm]=\mp2J_\pm(2J_z\pm1)\label{com},
\end{equation}
we obtain
\begin{widetext}
\begin{align}
i\dot{\widetilde{U}}_{11}&=\frac{\alpha}{\sqrt{N}} J_+ J_-\widetilde{U}_{21},\\
i\dot{\widetilde{U}}_{21}&=\frac{\alpha}{\sqrt{N}}
\widetilde{U}_{11}+2 \Bigl[ \mu+ \Bigl(
\frac{\gamma}{\sqrt{N}}+\frac{g}{N}(1-\Delta)\Bigl)\Bigl(J_z-\frac{1}{2}\Bigl)\Bigl]\widetilde{U}_{21},\\
i\dot{\widetilde{U}}_{22}&=\frac{\alpha}{\sqrt{N}} J_- J_+\widetilde{U}_{12},\\
i\dot{\widetilde{U}}_{12}&=\frac{\alpha}{\sqrt{N}}
\widetilde{U}_{22}-2 \Bigl[ \mu+ \Bigl(
\frac{\gamma}{\sqrt{N}}+\frac{g}{N}(1-\Delta)\Bigl)\Bigl(J_z+\frac{1}{2}\Bigl)\Bigl]\widetilde{U}_{12}.
\end{align}
\end{widetext}

Now, the coefficients in front of the new operator variables
$\widetilde{U}_{ij}$ are diagonal in the common eigenbasis of $J^2$
and $J_z$, whence the standard method of solving systems of
differential equations can be easily applied. Combining the above
relations leads to the following second order homogeneous
differential equations for the operators $\widetilde{U}_{21}$ and
$\widetilde{U}_{12}$
\begin{align}
\ddot{\widetilde{U}}_{21}&+2 i
\Bigl[\mu+\Bigl(\frac{\gamma}{\sqrt{N}}+\frac{g}{N}(1-\Delta)\Bigl)\Bigl(J_z-\frac{1}{2}\Bigl)\Bigl]\dot{\widetilde{U}}_{21}\nonumber\\
&+\frac{\alpha^2}{N}J_+J_-\widetilde{U}_{21}=0,\\
\ddot{\widetilde{U}}_{12}&-2 i
\Bigl[\mu+\Bigl(\frac{\gamma}{\sqrt{N}}+\frac{g}{N}(1-\Delta)\Bigl)\Bigl(J_z+\frac{1}{2}\Bigl)\Bigl]\dot{\widetilde{U}}_{12}\nonumber\\
&+\frac{\alpha^2}{N}J_-J_+\widetilde{U}_{12}=0,
\end{align}
which admit the following solutions
\begin{align}
\widetilde{U}_{21}=2i C_1 \mathrm e^{-iF_1 t} \sin\Bigl(t \sqrt{M_1}\Bigl)\label{s1},\\
\widetilde{U}_{12}=2i C_2 \mathrm e^{-iF_2 t} \sin\Bigl(t
\sqrt{M_2}\Bigl)\label{s2}.
\end{align}
Here, $C_1$ and $C_2$ are some diagonal operators to be determined
and we have
\begin{align}
F_1&=\mu+\Bigl(\frac{\gamma}{\sqrt{N}}+\frac{g}{N}(1-\Delta)\Bigl)\Bigl(J_z-\frac{1}{2}\Bigl),\\
M_1&=F_1^2+\frac{\alpha^2}{N} J_+J_-,\\
F_2&=-\mu-\Bigl(\frac{\gamma}{\sqrt{N}}+\frac{g}{N}(1-\Delta)\Bigl)\Bigl(J_z+\frac{1}{2}\Bigl),\\
M_2&=F_2^2+\frac{\alpha^2}{N} J_-J_+.
\end{align}
Integrating the right-hand side of Eq.~(\ref{s1})  gives
\begin{align}
\widetilde{U}_{11}=-2 C_1 & \mathrm e^{-iF_1t} \frac{\sqrt{N M_1
}}{\alpha}\nonumber \\  &\times \Bigl[\cos\Big(t
\sqrt{M_1}\Bigl)+\frac{i F_1}{\sqrt{M_1}}\sin\Big(t
\sqrt{M_1}\Bigl)\Bigl]+ C_3.
\end{align}
The constant  operators $C_1$ and $C_3$ can be determined using  the
initial conditions (\ref{initial}) and the unitarity condition for
the time evolution operator, which yield $C_1=-\alpha/(2 \sqrt{N M_1
}){\bf 1}_B$ and $C_3=0$. Hence, we obtain
\begin{align}
U_{11}(t)&=\mathrm e^{-i G_1 t}\Bigl[\cos\Bigl(t \sqrt{M_1}\Bigl)+\frac{i
F_1}{\sqrt{M_1}}\sin\Bigl(t \sqrt{M_1}\Bigl)\Bigl]\label{u11},\\
U_{21}(t)&=-i J_- \frac{\alpha}{\sqrt{N M_1}} \mathrm e^{-i G_1 t}\sin\Bigl(t
\sqrt{M_1}\Bigl)\label{u21},
\end{align}
where
\begin{equation}
G_1=-\frac{\gamma}{2\sqrt{N}}+\frac{g}{2N}\Bigl[\Bigl(K+2 \Delta
J_z^2\Bigl)+2(1-\Delta)\Bigl(J_z-\frac{1}{2}\Bigl)\Bigl].
\end{equation}
Following the same method, we  find that

\begin{align}
U_{22}(t)&=\mathrm e^{-i G_2 t}\Bigl[\cos\Bigl(t \sqrt{M_2}\Bigl)+\frac{i
F_2}{\sqrt{M_2}}\sin\Bigl(t \sqrt{M_2}\Bigl)\Bigl]\label{u22},\\
U_{12}(t)&=-i J_+ \frac{\alpha}{\sqrt{N M_2}} \mathrm e^{-i G_2 t}\sin\Bigl(t
\sqrt{M_2}\Bigl)\label{u12},
\end{align}
where
\begin{equation}
G_2=-\frac{\gamma}{2\sqrt{N}}+\frac{g}{2N}\Bigl[\Bigl(K+2 \Delta
J_z^2\Bigl)+2(\Delta-1)\Bigl(J_z+\frac{1}{2}\Bigl)\Bigl].
\end{equation}
It is easy to see that the operators $G_1$ and $G_2$ are diagonal in the common basis of $J^2$ and $J_z$.
\subsection{Reduced density matrix}

Having determined the exact analytical form of the time evolution
operator, we are able to calculate the reduced density matrix of the
central spin. Indeed, from Eqs.~(\ref{ev}) and~(\ref{tr}), and by
making use of the trace properties of the lowering and raising
operators $J_\pm$, we find that
\begin{align}
\rho_{11}(t)&=\frac{1}{Z_N}\Bigl[\rho_{11}^0 \mathrm {tr}_B\Bigl( \mathrm e^{-\beta
H_B} U_{11} U_{11}^*\Bigl)\nonumber \\&+\rho_{22}^0 \mathrm {tr}_B \Bigl(
\mathrm e^{-\beta H_B} U_{21}^*U_{12}\Bigl) \Bigl]\label{ev1},\\
\rho_{12}(t)&=\frac{1}{Z_N} \rho_{12}^0 \mathrm {tr}_B\Bigl( \mathrm e^{-\beta H_B}
U_{11} U_{22}^*\Bigl)\label{ev2}.
\end{align}
Furthermore, with the help of the commutation relations (\ref{com}), we can easily prove  that
 $ J_- F_1=-F_2 J_-$, and $J_- M_1=M_2 J_-$. Using the latter equalities, one can check
that the time-dependent components of the Bloch vector are given  by
\begin{widetext}
\begin{align}
\lambda_3(t)&=-\frac{2}{Z_N} \mathrm {tr}_B\Bigl\{\frac{\alpha^2 J_+J_-}{N M_1}
\mathrm e^{-\frac{g\beta}{2N} [K+2 \Delta J_z^2+ (1-\Delta)(2J_z-1)]}
\sin^2\Bigl( t \sqrt{M_1}\Bigl) \sinh\Bigl[\frac{g
\beta}{2N}(1-\Delta)(2J_z-1)\Bigl]\Bigl\}\nonumber\\
&+\lambda_3(0) \Bigl\{1-\frac{2}{Z_N} \mathrm {tr}_B\Bigl\{\frac{\alpha^2 J_+J_-}{N
M_1} \mathrm e^{-\frac{g\beta}{2N} [K+2 \Delta J_z^2+ (1-\Delta)(2J_z-1)]}
\sin^2\Bigl( t \sqrt{M_1}\Bigl) \cosh\Bigl[\frac{g
\beta}{2N}(1-\Delta)(2J_z-1)\Bigl]\Bigl\}\Bigl\},\label{lam3}\\
\lambda_{1}(t)&=\mathrm {tr}_B\Bigl\{\Bigl(\lambda_1(0)\cos(\Omega
t)+\lambda_2 (0)\sin(\Omega t)\Bigl) A
-\Bigl(\lambda_1(0)\sin(\Omega t)-\lambda_2(0) \cos(\Omega t)\Bigl) B\Bigl\}, \label{lam1}\\
\lambda_2(t)&=-\mathrm {tr}_B\Bigl\{\Bigl(\lambda_1(0)\sin(\Omega
t)-\lambda_2(0) \cos(\Omega t)\Bigl) A
+\Bigl(\lambda_1(0)\cos(\Omega t)+\lambda_2(0) \sin(\Omega t)\Bigl)
B \Bigl\},\label{lam2}
\end{align}
where
\begin{align}
\Omega &=\frac{2 g}{N}(\Delta-1) J_z,\\
 A&=\frac{1}{Z_N}
\Bigl\{\mathrm e^{-\frac{g\beta}{2N} [K+2 \Delta
J_z^2]}\Bigl[\cos\Bigl(t \sqrt{M_1}\Bigl) \cos\Bigl(t
\sqrt{M_2}\Bigl)+\frac{F_1F_2}{\sqrt{M_1M_2}} \sin\Bigl(t
\sqrt{M_1}\Bigl)\sin\Bigl(t \sqrt{M_2}\Bigl)\Bigl]\Bigl\},\\
B&=\frac{1}{Z_N}\Bigl\{\mathrm e^{-\frac{g\beta}{2N} [K+2 \Delta
J_z^2]}\Bigl[\frac{F_1}{\sqrt{M_1}} \sin\Bigl(t
\sqrt{M_1}\Bigl)\cos\Bigl(t \sqrt{M_2}\Bigl)-\frac{F_2}{\sqrt{M_2}}
\sin\Bigl(t \sqrt{M_2}\Bigl)\cos\Bigl(t
\sqrt{M_1}\Bigl)\Bigl]\Bigl\}.
\end{align}

\end{widetext}
\begin{figure}[htba]
{\centering
\resizebox*{0.45\textwidth}{!}{\includegraphics{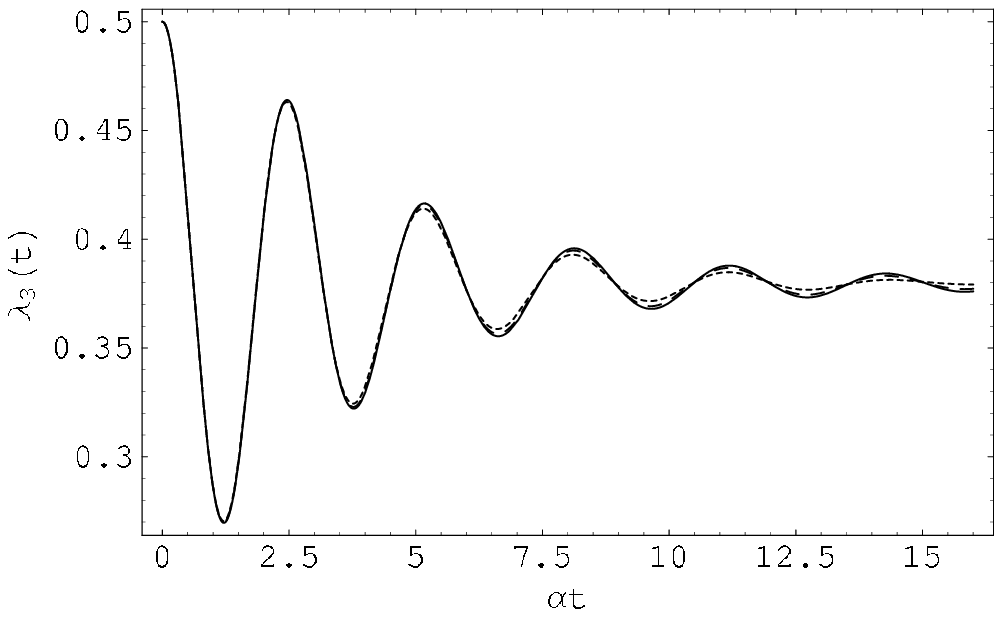}}
\resizebox*{0.45\textwidth}{!}{\includegraphics{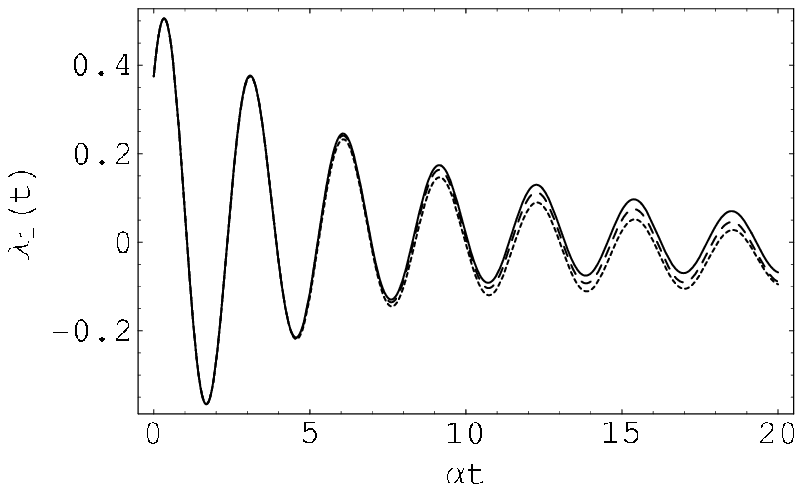}}
\par}
\caption{\label{figure1}  Time evolution of the components $\lambda_3(t)$ and $\lambda_1(t)$
for different values of the number of spins in the environment: $N=100$ (dotted lines), $N=200$
(dashed lines), and $N=400$ (solid lines).
The other parameters are $\gamma=0$, $g=1$, $\beta=0.5$, $\Delta=0$, and $\mu = \alpha$.
The
initial conditions are $\lambda_3(0)=\frac{1}{2}$, $\lambda_{1,2}(0)=\frac{3}{8}$.}
\end{figure}
 \begin{figure}[t]
{\centering\subfigure[{}]{
\resizebox*{0.45\textwidth}{!}{\includegraphics{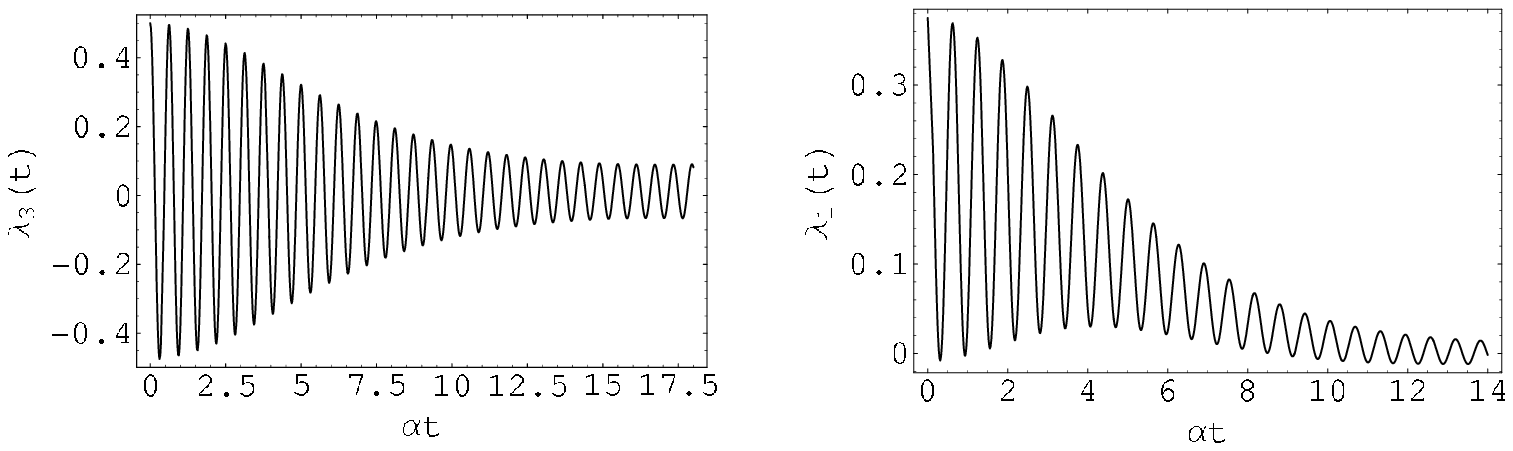}}}
\subfigure[{}]{
\resizebox*{0.45\textwidth}{!}{\includegraphics{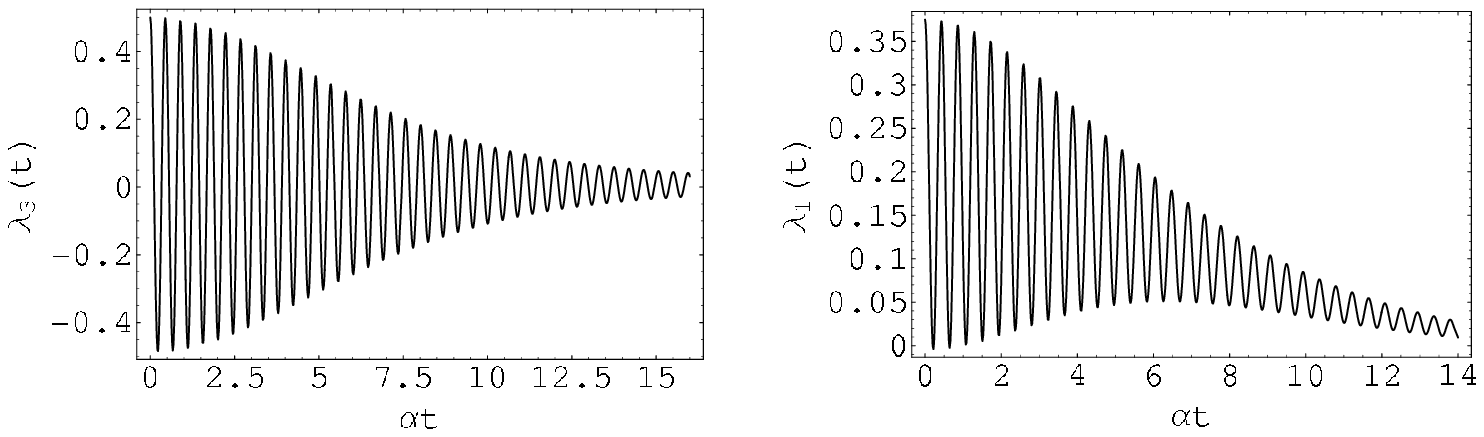}}}
\par}
\caption{\label{figure2} The Gaussian decay of the Bloch vector
components $\lambda_3(t)$ and $\lambda_1(t)$ in the case of
ferromagnetic interactions:
  (a) N=100 and (b) N=200. The plot on the left of each subfigure corresponds to $\lambda_{3}(t)$, the one on the right corresponds to
  $\lambda_1(t)$. The other parameters are
 $\gamma = 2\alpha$, $\beta  =1$, $g=-5$, $\Delta = 0.5$, $\mu = 0$, $\lambda_3(0)=\frac{1}{2}$,
 and $\lambda_{1,2}(0)=\frac{3}{8}$. }
\end{figure}

 From here on, the parameters $\mu$ and $\gamma$
will be given in units of the coupling
constant $\alpha$. The behavior of the component $\lambda_2(t)$ does not significantly
differ from the one corresponding to $\lambda_1(t)$.
Throughout the remainder of the paper we shall deal with the latter component and
restrict ourselves to positive values of the
anisotropy constant $\Delta$.

 Depending on the nature of  interactions  within the bath, we can distinguish two different cases.
  The first one corresponds to  positive values of $g$, i.e.   antiferromagnetic couplings between
  the  constituents of the environment. In this case,  as the number of spins  increases, the plots saturate and
   a nontrivial
  limit exists as shown in Fig.~\ref{figure1}. This will be investigated in the following section.
   The other case corresponds to negative values of  $g$, i.e.  ferromagnetic couplings within the bath.  When $\Delta<1$
   the components of the Bloch
   vector  exhibit in general  Gaussian  decay accompanied by fast damped oscillations even when the strength of
   the magnetic field is very weak. In contrast to $\lambda_1(t)$, the
   component $\lambda_3(t)$ decays faster as the number of spins increases. When the latter is small, $\lambda_3(t)$ may
   revive to decay again and so forth.  The numerical simulation shows that the details of  the time evolution
    of the reduced density matrix
   are rather complex and depend on the different values
    of the parameters of the model, including the number of bath spins.
   For example if we set
   $\Delta=0$, we observe that the oscillations are quickly suppressed with the increase of the strength of the magnetic
    field,  or the value of the coupling constant $\gamma$. In this case, the
   components $\lambda_{2,3}(t)$ do not vanish at long time scales; the corresponding asymptotic values depend, however, on $N$
   in contrast to the antiferromagnetic case. For large  values
   of the coupling constant $g$,  the component  $\lambda_1(t)$  quickly decays whereas  $\lambda_{3}(t)$  oscillates around
    zero with large amplitudes (typically of the same  order of magnitude as the corresponding initial value).
    We also notice that the frequencies of the damped oscillations increase with the increase of
    the number of bath spins as shown in Fig.~\ref{figure2}.
      Roughly speaking, when $\Delta>1$, the  behavior of the components of the
    Bloch vector   is quiet similar to  the antiferromagnetic counterpart. For example when $\gamma=0$, the
    components $\lambda_i(t)$ show saturation behavior with respect to the number
    of spins $N$; their asymptotic values are different from zero.

 In order to explain the differences
    between the behavior
    of the reduced density matrix in the ferromagnetic and the antiferromagnetic environments,
     we note that in the latter case,
   the form of interactions favors antiparallel spins. This is the reason for which the ground
    state of the antiferromagnetic bath, $|\Psi_G\rangle$,  is equal
    to $|0,0\rangle$.  On the
   contrary,  ferromagnetic interactions  force the spins in the bath to align along an arbitrary
    direction in the space.  In this case $|\Psi_G\rangle$   belongs to the subspace $\mathbb C^{N+1}$
     spanned by the state vectors
    $|\frac{N}{2},m\rangle$  corresponding
    to $j=\frac{N}{2}$ . For
    instance, when $\Delta>1$,  the ground state  of
     the bath turns out to be doubly degenerate, namely $|\Psi_G\rangle=|\frac{N}{2},\pm \frac{N}{2}\rangle$.
     For $\Delta<1$, we simply have
     $|\Psi_G\rangle=|\frac{N}{2},0\rangle$. However, when
    $\Delta=1$,
    the  ground energy of the bath is independent of the quantum number $m$; the degeneracy
    of $|\Psi_G\rangle$ is equal to $N+1$. Hence we conclude that the Hamiltonian $H_B$ displays
    quantum phase transition at $\Delta=1$. This is the reason for which
    the reduced dynamics  depends on whether the anisotropy constant is less or greater than one.
     Note that the mean value of $J^2$ is close to zero
     in the case of antiferromagnetic interactions within the spin bath  in contrast with
      the ferromagnetic case where $\langle J^2\rangle\sim N^2$. Obviously,
   the central spin   decoheres less if the spin bath, to which it couples, is characterized by a total angular
    momentum close to zero.
   At zero
   temperature, the antiferromagnetic bath  occupies its ground state $|0,0\rangle$ which is an eigenvector of $H_B$,
   and satisfies $H_{SB}|\pm\rangle\otimes|0,0\rangle=0$.
   Hence, the central spin
  remains  decoupled from the bath if the initial state factorizes: the two-level system
    preserves its coherence regardless of
    the number of environmental spins. Let us now consider the case where $\gamma=0$ and  $\Delta>1$.
      At low temperatures, the total angular momentum of the ferromagnetic bath
     has the tendency
   to be directed along the $z$ direction. Since the central spin couples to the bath through
    Heisenberg $XY$ interactions ($\gamma=0$), we end up with
   a situation quiet similar to  that where $g>0$. The above results show that properties of the bath at zero temperature affect the behavior
   of the reduced dynamics when $T>0$.
   At infinite temperature,  the ferromagnetic and antiferromagnetic environments become completely unpolarized;
   the reduced dynamics displays the same behavior
   in both systems as $N$ increases.

\section{The limit  $N \to\infty$}

 This section is devoted to  the case of an infinite number of spins in the environment, i.e. the case
$N\to\infty$. We investigate the effect of the bath temperature, the external magnetic field,  and the anisotropy
constant on the reduced density matrix of the central spin.
 To this end it should be noted that the trace of  the
 operators $J_\pm/\sqrt{N}$  together with
$J_z/\sqrt{N}$ is identically zero, namely
\begin{equation}
\mathrm {tr}_B\Bigl\{\frac{J_\pm}{\sqrt{N}}\Bigl\}=\mathrm {tr}_B\Bigl\{\frac{J_z}{\sqrt{N}}\Bigl\}=0.
\end{equation}
A  more general property of the trace of the lowering and raising operators
can be  expressed as
\begin{align}
\lim\limits_{N\to\infty}& 2^{-N} \mathrm
{tr}_B\Bigl\{\prod\limits_{i=1}^{k} \Bigl(\frac{J_\pm
J_\mp}{N}\Bigl)^{n_i}\Bigl\}\nonumber \\ &=\lim\limits_{N\to\infty}
2^{-N} \mathrm {tr}_B\Bigl\{\prod\limits_{i=1}^{k}
\Bigl(\frac{J_\pm}{\sqrt{N}}\Bigl)^{n_i}\Bigl(\frac{
J_\mp}{\sqrt{N}}\Bigl)^{n_i}\Bigl\}=\frac{n!}{2^n}\label{gauss},
\end{align}
where $n=\sum\limits_{i=1}^{k} n_i$ is  positive integer;  the
 trace vanishes for all the cases in which $J_+$ and $J_-$ appear
with different exponents. This means that $J_{\pm}/\sqrt{N}$
 are well-behaved fluctuation operators with respect to the tracial state. Hence  in the limit $N\to\infty$, the
 operator $J_+/\sqrt{N}$
 converges to a  complex random variable $z$ with the probability density
 function~\cite{16}
 \begin{equation}
 z\mapsto\frac{2}{\pi} \mathrm e^   {-2 |z|^2}.\label{zz}
\end{equation}
Here, we wish to mention the similarity that exists between relation~(\ref{gauss}) and
\begin{equation}4\int\limits_{0}^{\infty} \ t \   dt \  t^{2n} \mathrm e^{-2 t^2} =
\frac{n!}{2^{n}}\end{equation} which is a special case of $
\int\limits_{0}^{\infty} t^{2n+1} \mathrm e^{-a t^2} dt = \frac{n!}{2
a^{n+1}}$, where $n=0,1,2,...$, and the real part of $a$ satisfies $\mathrm{Re}(a)>0$.

The operator   $J_z/\sqrt{N}$ also converges to a real random
variable  $m$ (to be differentiated from the eigenvalue $m$) when
$N\to\infty$,  with the probability density function
\begin{equation}
m\mapsto \sqrt{\frac{2}{\pi}} \mathrm e^   {-2 m^2}.\label{m}
\end{equation}
 For example, consider  the operator $\mathrm e^{-i\frac{2 \gamma t}{\sqrt{N}} J_z}$ and let us calculate
\begin{equation}
\mathrm {tr}_B\Bigl\{\mathrm e^{-i \frac{2\gamma t}{\sqrt{N} }J_z}\Bigl\}=\prod\limits_{k=1}^{N}\mathrm{tr}
\mathrm \ e ^{-i\frac{\gamma t}{\sqrt{N}}\sigma_z^k}.
\end{equation}
The trace under the product in the right-hand side of the above equation can be easily evaluated as
$2 \cos(\frac{\gamma t}{\sqrt{N}})$. Consequently,
\begin{equation}
\mathrm {tr}_B\Bigl\{\mathrm e^{-i \frac{2\gamma t}{\sqrt{N} }J_z}\Bigl\}=2^N \Bigl[\cos\Bigl(\frac{\gamma t}{\sqrt{N}}\Bigl)\Bigl]^N.
\end{equation}
Expanding the cosine function in a Taylor series and taking the limit $N\to\infty$ yield
\begin{align}
\lim\limits_{N\to\infty}2^{-N}\mathrm {tr}_B\Bigl\{\mathrm e^{-i \frac{2\gamma t}{\sqrt{N}} J_z}\Bigl\}&=\lim\limits_{
N\to\infty}\Bigl[1-\frac{\gamma^2 t^2}{2N}+O\Bigl(\frac{1}{N^2}\Bigl)\Bigl]^N\nonumber\\&=\mathrm e^{-\frac{\gamma^2 t^2}{2}}
\label{gamma}.\end{align}
On the other hand we have
\begin{equation}
\sqrt{\frac{2}{\pi}}\int\limits_{-\infty}^{\infty} \mathrm e^{-2 m^2 -2i \gamma t m} dm=\mathrm e^{-\frac{\gamma^2 t^2}{2}},\label{hoo}
\end{equation}
which is in agreement with Eq.~(\ref{gamma}). In particular we can infer that
\begin{equation}\lim\limits_{N\to\infty}2^{-N} \mathrm{tr}_B \Bigl(J_z/\sqrt{N}\Bigl)^{2n}=
\frac{\Gamma(n+\frac{1}{2})}{2^n\sqrt{\pi}},
\end{equation} where $\Gamma(z)$ is Euler gamma function.
 We shall use the latter results when we investigate the short-time behavior
of the reduced density matrix in the case where $\gamma$ is different from zero.

One can check that for large values of $N$,
\begin{align}
\mathrm{tr}_B\Bigl\{\Bigl(\frac{J_z}{\sqrt{N}}\Bigl)^{k}\Bigl(\frac{J_\pm J_\mp}{N}\Bigl)^\ell\Bigl\} \approx & 2^{-N}
 \mathrm{tr}_B\Bigl\{\Bigl(\frac{J_z}{\sqrt{N}}\Bigl)^{k}\Bigl\}\nonumber\\ & \times \mathrm{tr}_B\Bigl\{\Bigl(\frac{J_\pm J_\mp}{N}\Bigl)^\ell\Bigl\}.
\label{uncor}\end{align}
 For odd powers of $J_z$, the left-hand side of the above relation  vanishes
  as $N $ increases; the right-hand side is always zero. Eq.~(\ref{uncor}) simply implies that the operators
   $J_\pm J_\mp/N$ and $J_z/\sqrt{N}$ become uncorrelated under  the tracial state at large values of $N$.
 Note that the above state  corresponds to a bath of $N$ independent spin-$\frac{1}{2}$ particles, i.e. the
 state of maximum entropy.
 In the limit of large number of spins such a bath has the tendency to behave as a classical stochastic system.
 The scaled bath
 operators $J_\alpha/\sqrt{N}$ (where $\alpha\equiv x, y,z$) converge to independent commuting random variables.
  For instance, we can easily show that the trace over the environmental degrees of freedom of the
  operator $\exp\Bigl[\frac{\epsilon t}{\sqrt{N}}\Bigl(a_1 J_x+a_2J_y+a_3J_z\Bigl)\Bigl]$,
   where $a_{1,2, 3}\in \mathbb C$
 and $\epsilon=\sqrt{\pm1}$, is given by
   $2^N \Bigl\{\cosh\Bigl[\frac{\epsilon t}{2\sqrt{N}}\sqrt{a^2_1+a^2_2+a^2_3}\Bigl]\Bigl\}^N$. If we expand  the $\cosh$ function in
   Taylor series and take the limit $N\to\infty$,  as we did in Eq.~(\ref{gamma}), we end up with the
   result
     $\exp[\frac{\epsilon^2 t^2}{8}(a^2_1
   +a^2_2+a^2_3)]$. The latter can be obtained by multiple integration over three independent random variables each of which has
    the same probability density function as $m$ [see Eq.~(\ref{hoo})].  It follows that the
  random variables  $z$ and $m$ can be treated as independent
  in the limit $N\to\infty$.

    From the above discussion, we can conclude that
\begin{widetext}
\begin{equation}
\lim\limits_{N\to\infty}2^{-N}\mathrm {tr}_B\Bigl\{f\Bigl(\frac{J_\pm
J_\mp}{N},\frac{J_z}{\sqrt{N}}\Bigl)\Bigl\}=\Bigl(\frac{2}{\pi}\Bigl)^{3/2}\int\limits_{-\infty}^{\infty}
dm \int\limits_{\mathbb{C}} dz dz^*f(|z|^2,m)\  \mathrm e^   {-2 (m^2+|z|^2)}\label{lim}
\end{equation}
\end{widetext}
at least for bounded functions $f: \mathbb C \times \mathbb{R} \to
\mathbb{R}$. The latter relation has been numerically checked for
large number of functions; the agreement between its two sides is
perfect. In fact, the class of functions for which the integral in
the right-hand side of Eq.~(\ref{lim}) exists contains  all the
functions having the form $\mathrm e^{-(a |z|^2+ b m^2)}h(|z|^2,m)$
where $h$ is bounded and $a$ and $b$ are complex numbers satisfying
$\mathrm{Re}(a)> -2, \mathrm{Re}(b)> -2$. If the latter conditions
are not satisfied then the integral does not converge. This is the
reason for which we shall restrict ourselves to the
antiferromagnetic case where  $g$ and $\Delta$ are positive.

  Under the above assumptions,
 it is possible
to evaluate the quantity
\begin{align}
\bar{Z}&=\lim\limits_{N\to\infty}2^{-N}Z_N\nonumber\\
&=\Bigl(\frac{2}{\pi}\Bigl)^{3/2}\int\limits_{-\infty}^{\infty} dm
\int\limits_{\mathbb{C}} dz dz^* \mathrm e^   {-(2+g\beta\Delta)
m^2-(2+g\beta)|z|^2}
\end{align}
 by making use of the polar
coordinates $(r, \phi)$ where $z= r \mathrm e^{i \phi}$. A straightforward calculation yields
\begin{equation}
\bar{Z}=\frac{2\sqrt{2}}{(2+g\beta)\sqrt{2+g\beta\Delta}}\to
\frac{2}{(2+g\beta)}\label{z}
\end{equation}
when ${\Delta\to 0}$. Obviously, if $g \beta=0$ then $\bar{Z}=1$. The agreement
between the right-hand side and the left-hand side of Eqs.~(\ref{lim}) is illustrated in Table~\ref{ZN} where we display
  $2^{-N} Z_N$ at different values of  $N$ and compare it with $\bar{Z}$ for $g=2$, $\Delta=5$ and $\beta=1$; the agreement is clearly very good for $N=5000$.

\begin{table}
\caption{\label{ZN} $2^{-N}Z_N$ at different values of $N$ for
$g=2$, $\Delta=5 $ and $\beta=1$; $\bar{Z}$=0.204124.}
\begin{ruledtabular}
\begin{tabular}{|c|c|c|c|c|}
$N$ & 10 & 100 & 1000 & 5000\\
\colrule
$2^{-N}Z_N$ & 0.203026  & 0.203997  & 0.204111& 0.204122 \\
\end{tabular}
\end{ruledtabular}
\end{table}
\begin{figure}[htba]
{\centering
\resizebox*{0.40\textwidth}{!}{\includegraphics{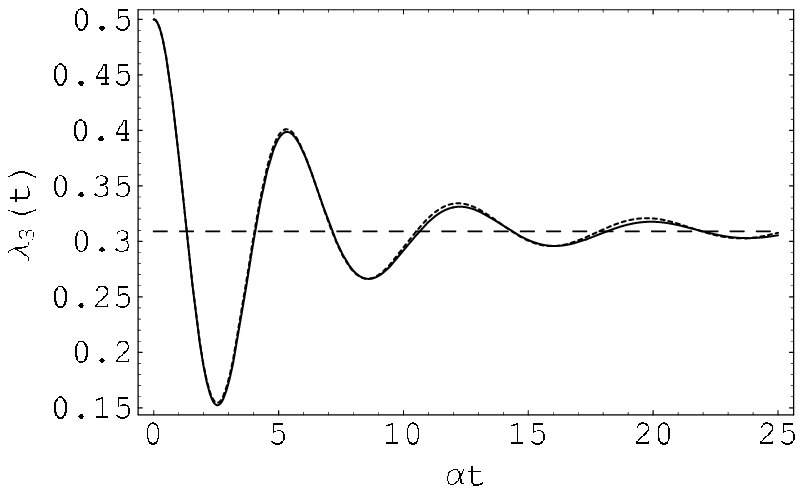}}
\resizebox*{0.45\textwidth}{!}{\includegraphics{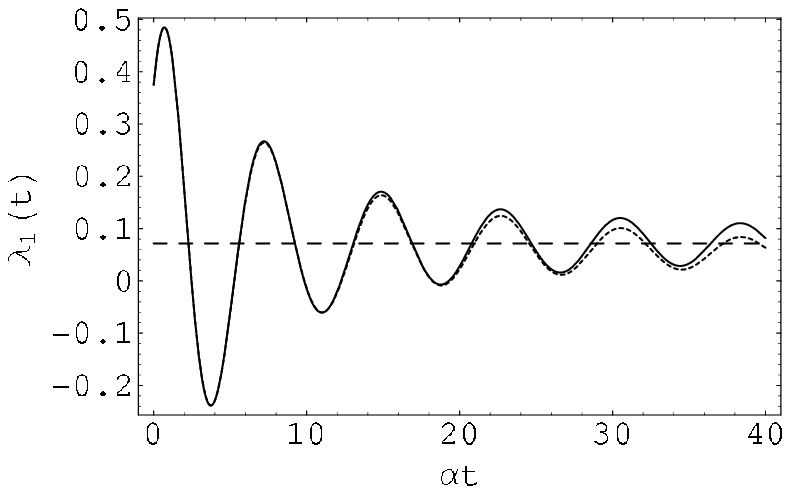}}
\par}

\caption{  \label{figure3}Evolution in time of $\lambda_3(t)$ and
$\lambda_1(t)$  for $N=400$ (dotted lines), and  $N\to\infty$ (solid
lines); the  dashed lines correspond to the asymptotic values.
 Other parameters are  $\gamma=0$, $g=1$, $\beta=5$, $\Delta=0.5$, $\mu = 0.4\alpha$, $\lambda_3(0)=\frac{1}{2}$ and $\lambda_{1,2}(0)=\frac{3}{8}$.}
\end{figure}

 Let us now focus on the general structure of
Eqs.~(\ref{lam3})-(\ref{lam2}). Clearly, we need to evaluate terms
having the general form

\begin{equation}
\frac{1}{Z_N} \mathrm {tr}_B\Bigl\{f\Bigl(\frac{J_\pm J_\mp}{N},\frac{J_z}{\sqrt{N}}\Bigl)\Bigl\}=\frac{
2^{-N}\mathrm {tr}_B\Bigl\{f\Bigl(\frac{J_\pm J_\mp}{N},\frac{J_z}{\sqrt{N}}\Bigl)\Bigl\}}{2^{-N} Z_N}.
\end{equation}
 As $N\to\infty$, the previous quantity tends to
\begin{equation}
\langle f\rangle =\bar{Z}^{-1}4 \Bigl
(\frac{2}{\pi}\Bigl)^{1/2}\int\limits_{-\infty}^{\infty} dm
\int\limits_{0}^{\infty} r dr f(r^2,m) \ \mathrm e^   {-2 (m^2+r^2)}\label{tran}.
\end{equation}
This is permissible since  the functions of interest appearing in
Eqs~(\ref{lam3})-(\ref{lam2}) fulfil all the conditions mentioned
above.  Note that the factor 4  in Eq.~(\ref{tran}) appears after
performing the
 integration with respect to the polar coordinate $\phi$ (
this actually  follows from the  symmetry with respect to the $z$
direction). It is also  quiet interesting to notice that
  the behavior of the central spin when  $-2< g\beta <0$ and $-2< g\beta \Delta<0$
is similar to that where $g>0$ and $\Delta>0$  as indicated by the conditions on the convergence of the integral in Eq.~(\ref{lim}).

\subsection{The case  $\gamma=0$}
Let us assume that the coupling constant $\gamma$ is
equal to zero. First of all, it should be noted that, although  the operator $J_z/\sqrt{N} $ converges to a random
variable,
 we can neglect the contribution of  $J_z/N $ when $N$ becomes very large.
  This means that in the limit $N\to\infty$, the quantities $M_{1,2}$ do not
 depend on the random variable $m$;  the $\sinh$ ($\sin$)
and the $\cosh$ ($\cos$) functions appearing in  Eq.~(\ref{lam3}) [Eqs.~(\ref{lam1})-(\ref{lam2})] should be
replaced  by zero and one, respectively. We only need  to integrate with respect to the random variable $z$ since the
integrals with respect to $m$ occurring in the numerator and denominator of Eq.~(\ref{tran}) cancel each other. One then
 concludes that
  the anisotropy
constant $\Delta$ has no effect on the dynamics of the  central spin when $N\to\infty$. This is  due to the fact that $H_{SB}$ simplifies
to Heisenberg $XY$ Hamiltonian. Only transverse interactions contribute to the reduced dynamics when $N$ is sufficiently
 large because  $J_z/\sqrt{N}$ and $K/N$ (or equivalently $J_\pm J_\mp/N$) become practically uncorrelated under the tracial state [see Eq.(\ref{uncor})].

 Hence,  in the limit of an infinite number of spins
within the bath we obtain
\begin{equation}
\lambda_3(t)=\lambda_3(0) \Bigl (1- \eta(t)\Bigl),\end{equation} where
\begin{equation}
\eta(t)=\Bigl\langle 2 r^2
\frac{\sin^2\Bigl( t\sqrt{\mu^2+ r^2}\Bigl)}{\mu^2+
r^2} \mathrm e^{-g\beta[r^2+\Delta m^2]}\Bigl\rangle.\label{fun}\end{equation}
Note that the time variable is now given in units of $\alpha$. We show in the appendix that the above function
can be written as

\begin{widetext}
\begin{align}
 \eta(t)&=1-\cos\Bigl(2 \mu t\Bigl)+  \frac{i  t}{2} \sqrt{\frac{\pi}{2+g\beta}}
 \Bigl\{\text{erf}\Bigl[\frac{\mu(g\beta+2)-
 it}{\sqrt{g\beta+2}}\Bigl]-\text{erf}\Bigl[\frac{\mu(g\beta+2)+
 it}{\sqrt{g\beta+2}}\Bigl]\Bigl\}\ \mathrm e^{[(2+g\beta)
 \mu^2-\frac{t^2}{2+g\beta}]}\nonumber\\ &-(g\beta+2) \mu^2\ \mathrm e^{(g\beta+2)\mu^2}
 \Gamma\Bigl[0,(g\beta+2)\mu^2\Bigl]+\mu^2(g\beta+2)\ \mathrm e^{(g\beta+2)\mu^2} \mathrm{Re}\Bigl\{\Gamma\Bigl[0,(g\beta+2)\mu^2+2\mu i
 t\Bigl]\Bigl\}\nonumber \\&+\mu^2 \mathcal{M}(t;\mu,\beta),\label{eta}
 \end{align}
\end{widetext}
where
\begin{align}
\text{erf}(z)&=\frac{2}{\sqrt{\pi}}\int\limits_0^z e^{-t^2} dt ,\\
\Gamma(a,z)&=\int\limits_z^{\infty} t^{a-1} e^{-t} dt.
\end{align}
 are, respectively, the error and the incomplete gamma
functions~\cite{28}. The function $\mathcal{M}$ is given by Eq.~(\ref{M}) of the appendix.

The remaining components of the Bloch vector are given by
\begin{align}
\lambda_1(t)&=
\lambda_1(0)\Bigl[\zeta(t)+\frac{1}{2}\eta(t)\Bigl]+\lambda_2(0) \xi(t),\\
\lambda_2(t)&=
\lambda_2(0)\Bigl[\zeta(t)+\frac{1}{2}\eta(t)\Bigl]-\lambda_1(0) \xi(t),
\end{align}
where
\begin{widetext}
\begin{align}
\zeta(t)&=\Bigl\langle \mathrm e^{-g\beta (r^2+\Delta m^2)}\cos\Bigl(2 t \sqrt{\mu^2+
r^2}\Bigl)\Bigl\rangle\nonumber \\ &=\cos\Bigl(2 \mu t\Bigl)+\frac{i t}{2}
\ \mathrm e^{[(2+g\beta)
 \mu^2-\frac{t^2}{2+g\beta}]}\sqrt{\frac{\pi}{2+g\beta}}
 \Bigl\{\text{erf}\Bigl[\frac{\mu(g\beta+2)+
it}{\sqrt{g\beta+2}}\Bigl]-\text{erf}\Bigl[\frac{\mu(g\beta+2)-
 it}{\sqrt{g\beta+2}}\Bigl]\Bigl\}\label{zeta},
\end{align}
and
\begin{align}
\xi(t)&=\Bigl\langle  \mu\  \mathrm e^{-g\beta (r^2+\Delta m^2)} \frac{\sin\Bigl(2 t \sqrt{\mu^2+
r^2}\Bigl)}{\sqrt{\mu^2+r^2}}\Big\rangle \nonumber\\&=\frac{i\mu}{2} \sqrt{\pi
(2+g\beta)}\ \mathrm e^{[(2+g\beta)
 \mu^2-\frac{t^2}{2+g\beta}]}\Bigl\{\text{erf}\Bigl[\frac{\mu(g\beta+2)-
 it}{\sqrt{g\beta+2}}\Bigl]-\text{erf}\Bigl[\frac{\mu(g\beta+2)+
 it}{\sqrt{g\beta+2}}\Bigl]\Bigl\}.\label{xi}
\end{align}
\end{widetext}

The asymptotic behavior of the  reduced density matrix can be easily
determined as follows. Let us begin with the simplest functions
namely $\zeta(t)$ and $\xi(t)$. Their limits when $t\to\infty$ are
equal to zero which immediately follows from the Riemann-Lebesgue lemma
\begin{equation}
\lim\limits_{t\to\infty} \zeta(t)=\lim\limits_{t\to\infty} \xi(t)=0.
\end{equation}
The same lemma can be applied to the function $\eta(t)$ after some
simplifications of the integrals of interest as shown in the
appendix. Only one term survives the above approach when $t$ goes to
infinity, namely
\begin{equation}
\lim\limits_{t\to\infty}\eta(t)=1-\eta^{\infty},\label{et}
\end{equation}
where
\begin{equation} \eta^{\infty}=\mu^2(g\beta+2)\
\mathrm e^{\mu^2(g\beta+2)}\Gamma\Bigl(0,\mu^2(g\beta+2)\Bigl).\label{assym}
\end{equation}
 Hence, the asymptotic behavior of the reduced density matrix can be expressed  as
\begin{equation}
\lim\limits_{t\to\infty}\vec{\lambda}(t)=\vec{\lambda}_0-\eta^{\infty}\mathcal{W}\vec{\lambda}(0)\label{blochas}
\end{equation}
with
\begin{equation}
\vec{\lambda}_0=\frac{1}{2}\begin{pmatrix}\lambda_1(0)\\ \lambda_2(0)\\0\end{pmatrix},
\quad\mathcal{W}=\begin{pmatrix}\frac{1}{2}&0&0\\0&\frac{1}{2}&0\\0&0&-1\end{pmatrix}.
\end{equation}
 The evolution in time of
the components $\lambda_3(t)$ and $\lambda_1(t)$ is shown in Fig.~\ref{figure3} for $N=400$ spins in the environment,
along with the corresponding infinite case and the asymptotic limits obtained in Eq.~(\ref{blochas}).
We  can see that the off-diagonal elements of the
reduced density matrix show partial decoherence.  At low temperature, the relevant bath states
  are those
 with low energies (i.e. $j$ close to zero). In this case, the central spin is weakly coupled to
 the bath and hence preserves most of its coherence. At high temperature, the two-level system becomes more correlated
  with the bath which, however, behaves
 as a system of independent uncoupled particles. Thus quantum fluctuations
 within the antiferromagnetic spin-environment reduce  the effect of the decoherence of the central spin.
  In the following, we
 discuss how  the bath temperature and the strength of the
applied magnetic field affect the decay of the elements of the reduced density matrix.
  \begin{figure}[htba]
{\centering
\resizebox*{0.45\textwidth}{!}{\includegraphics{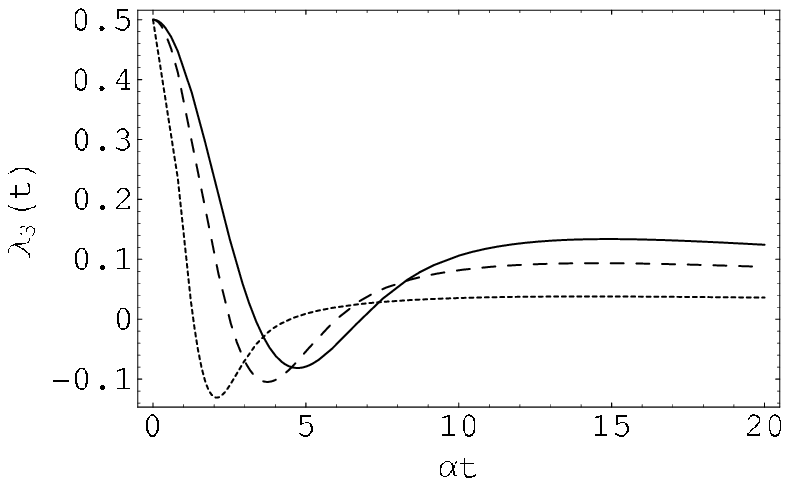}}
\resizebox*{0.45\textwidth}{!}{\includegraphics{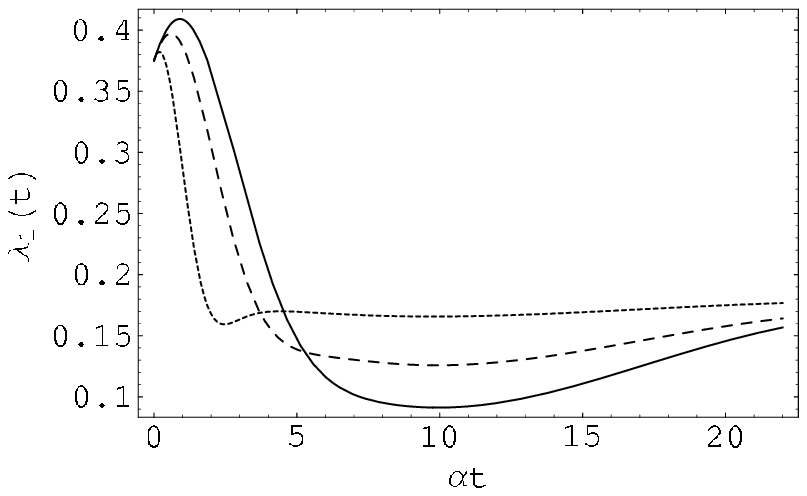}}
\par}

\caption{  \label{figure100}The decay of  the
components$\lambda_3(t)$ and $\lambda_1(t)$  for  $g=0$ (dotted
lines), $g=5$ (dashed lines), and $g=10$ (solid lines).
 Other parameters are  $\beta=1$, $\gamma=\Delta=0$, $\mu=0.1\alpha$, $\lambda_3(0)=\frac{1}{2}$ and $\lambda_{1,2}(0)=\frac{3}{8}$.}
\end{figure}

 Clearly, if $\mu=0$,
the vector component $\lambda_3(t)$ vanishes when $t\to\infty$
regardless of the bath temperature. This means that
$\rho_{11}(\infty)=\rho_{22}(\infty)=\frac {1}{2}$, which is
obviously independent of the initial state
 of the central system. On the contrary, the off-diagonal elements tend
  asymptotically to  half of their initial values. This follows from the fact that
 the temperature-dependent
quantity $\eta^{\infty}$  is proportional to the magnetic field
strength. The latter results are mainly due to the rotational
symmetry  of the model Hamiltonian together with the randomness of
the interactions within the bath.
  Fig~\ref{figure100} illustrates the difference in the decoherence process
   between the case of a static bath ($g=0$) and  a dynamic bath ($g\neq0$).
  We can see that the asymptotic value of
  $\lambda_1(t)$ decreases with the increase of $g$ in contrast to $\lambda_3(t)$
  which assumes larger asymptotic values when $g$ increases. However, at short times
  the above
 components decay slower with the increase of  $g$, implying that
strong quantum correlations within the
environment suppress the effect of the decoherence process~\cite{29}. As we shall see below, the decay of the reduced
density matrix exhibits a
  reverse behavior with respect to  the temperature of  the bath . This can be explained by the dependence of
 the decoherence time constant of our model, which turns out to be equal to
$\tau=\sqrt{\frac{2+g\beta}{\alpha^2}}$ as
 revealed by  Eqs.~(\ref{eta}), (\ref{zeta}) and (\ref{xi}),  on the product $g\beta$.
Clearly,  $\tau\to\infty$
  as $g\to\infty$ or/and $T\to0$, which confirms the above statements.
\begin{figure}[htba]
{\centering
\resizebox*{0.45\textwidth}{!}{\includegraphics{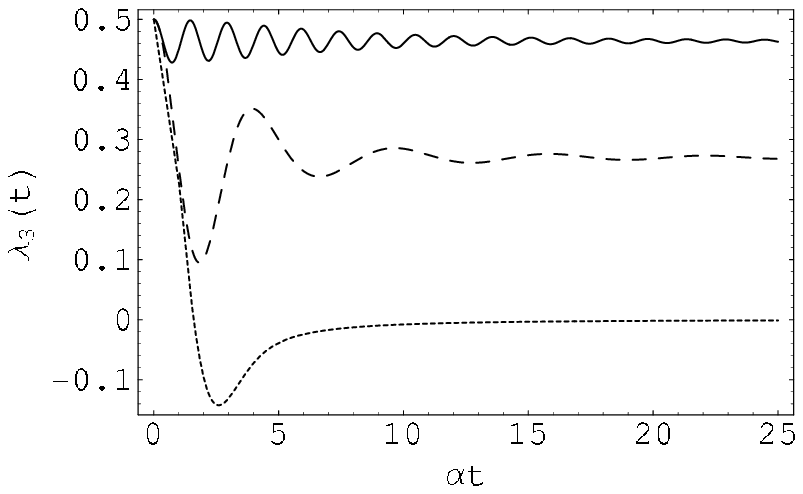}}
\resizebox*{0.45\textwidth}{!}{\includegraphics{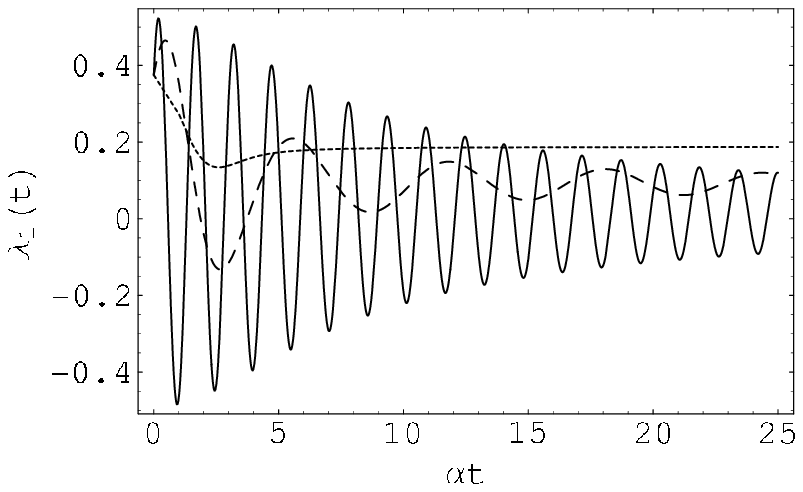}}
\par}

\caption{\label{figure4} Dependence of $\lambda_3(t)$ and $\lambda_1(t)$ on the strength of the
magnetic field in the case $N\to\infty$:  $\mu=0$ (dotted lines), $\mu=0.5\alpha$ (dashed lines), and $\mu=2\alpha$ (solid lines).
The parameters are  $\gamma=0$, $g\beta=2$, $\lambda_3(0)=\frac{1}{2}$ and $\lambda_{1,2}(0)=\frac{3}{8}$.}
\end{figure}

 Figure~\ref{figure4} illustrates the dependence of the components of the Bloch vector on the strength of the magnetic
field. We can see that  $\lambda_1(t)$  decays with the increase of $\mu$ whereas $\lambda_3(t)$
approaches its
initial value. Indeed, by making use of the following asymptotic expression of the incomplete
gamma function~\cite{28}
 \begin{equation}
 \Gamma(a,z)\sim z^{a-1}
 \mathrm e^{-z}\Bigl[1+\frac{a-1}{z}+\frac{(a-1)(a-2)}{z^2}+...\Bigl],
 \end{equation}
 when $z\to \infty$ in $|\arg z|< 3\pi/2$, we obtain
\begin{equation}
\lim\limits_{\mu,\beta \to\infty}\eta^{\infty}=1. \label{mil}
\end{equation}
Therefore, if  the ratio $\mu/\alpha$ is infinitely big then the off-diagonal elements of the reduced density matrix
 tend asymptotically to
 zero; the diagonal ones assume their initial
values. The above  results can be explained by the fact that  the effect of the bath
 on the dynamics of
 the central spin can be neglected when $\mu$ is very large compared to $\alpha$. The evolution in time
  is thus governed by the free Hamiltonian $H_S$ which does not
 affect the diagonal elements of the reduced density matrix.  The off-diagonal elements, however, show
 periodic oscillations; the vanishing asymptotic values obtained from Eqs.~(\ref{et}) and (\ref{mil}) will never
 be reached since the decoherence time constant is infinite ($\alpha\to0$).
\begin{figure}[htba]
{\centering
\resizebox*{0.45\textwidth}{!}{\includegraphics{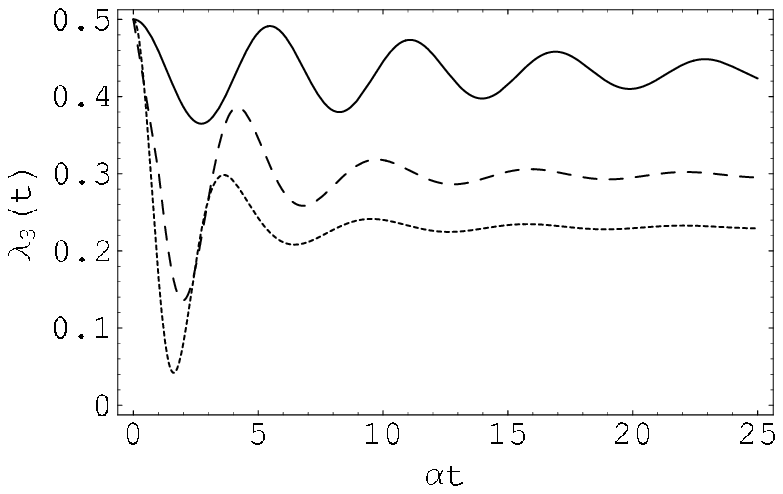}}
\resizebox*{0.45\textwidth}{!}{\includegraphics{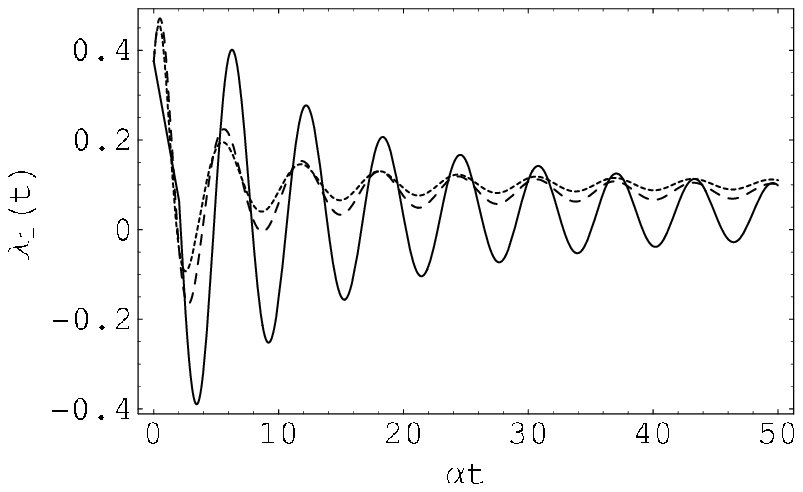}}
\par}

\caption{\label{figure9} Dependence of $\lambda_3(t)$ and $\lambda_1(t)$
 on the bath temperature in the case $N\to\infty$:  $\beta=0$ (dotted lines), $\beta=1$ (dashed lines), and $\beta=10$ (solid lines).
The parameters are  $\gamma=0$, $g=2$, $\mu=0.5 \alpha$, $\lambda_3(0)=\frac{1}{2}$ and $\lambda_{1,2}(0)=\frac{3}{8}$.}
\end{figure}

From Fig.~\ref{figure9} it can be seen that the bath temperature has a reverse effect on the decay of
the reduced density matrix elements.  The components $\lambda_3(t)$ and $\lambda_1(t)$ decay faster with the
 increase of  $T$. Furthermore, we can see that the diagonal elements  assume larger asymptotic values in contrast with the off-diagonal ones. In the limit
of zero temperature, the asymptotic behavior
is identical to the one corresponding to  $\mu\to\infty$, see  Eq.~(\ref{blochas}). Indeed, at zero
 temperature the bath and
the central spin evolve independently from each other  as we already mentioned in the previous section.
   Once again, we find that the dynamics of the central spin is governed by the free Hamiltonian $H_S$ which preserves
   the coherence of the central system.

\begin{figure}[htba]
{\centering
\resizebox*{0.45\textwidth}{!}{\includegraphics{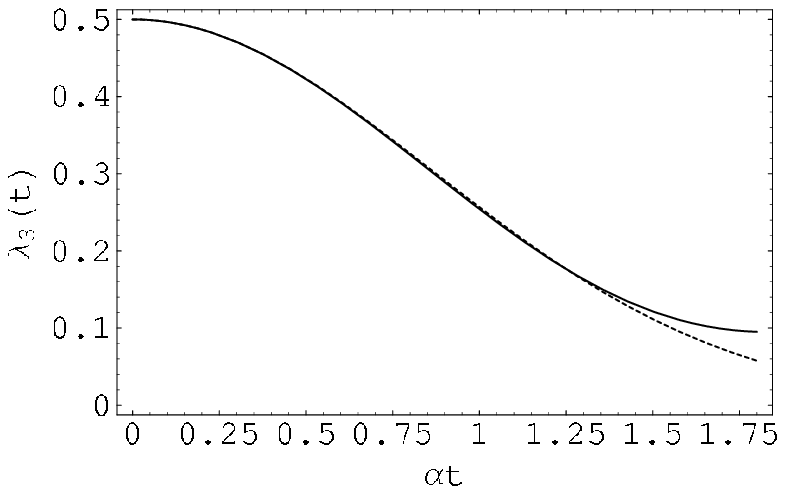}}
\resizebox*{0.45\textwidth}{!}{\includegraphics{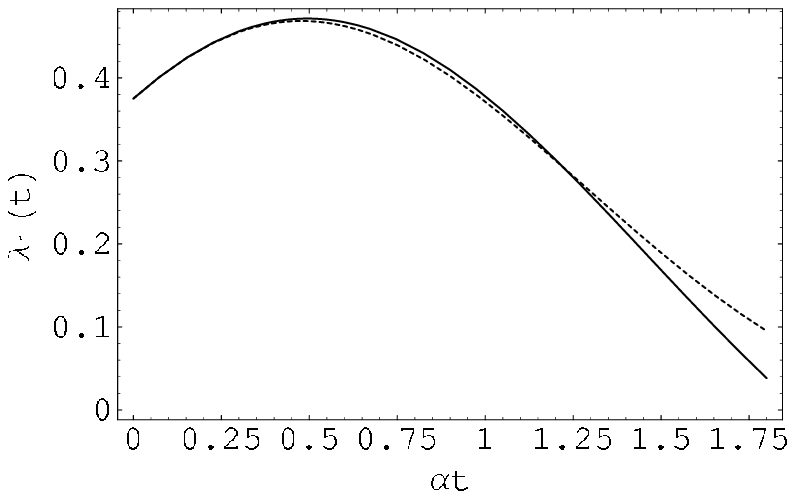}}
\par}
\caption{\label{figure11}  The short-time behavior of $\lambda_3(t)$ and $\lambda_1(t)$  in the case
$N\to\infty$. The solid lines correspond to the exact solutions, the dotted lines denote the approximations
 (\ref{short3}-\ref{short4}).
Here,  $\gamma=0$, $g\beta=1$, $\mu =0.5\alpha$, $\lambda_3(0)=\frac{1}{2}$ and $\lambda_{1,2}(0)=\frac{3}{8}$.}
\end{figure}
\begin{figure}[htba]
{\centering
\resizebox*{0.45\textwidth}{!}{\includegraphics{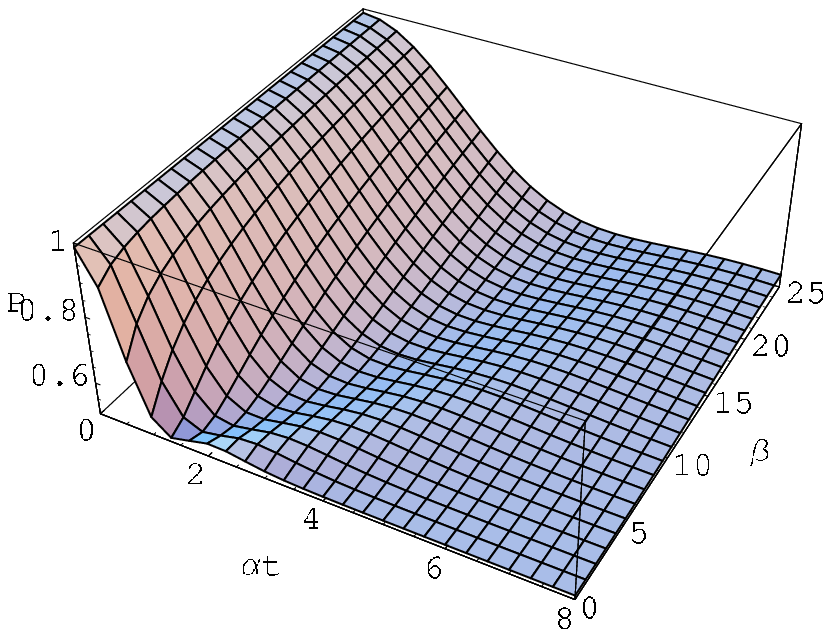}}
\par}
\caption{\label{figure13} (Color online) Purity evolution for different values of the bath temperature in the case
 $N\to\infty$ with
 $\gamma=0$, $g=1$ and $\mu=0$. The
initial conditions are $\lambda_3(0)=\sqrt{\frac{7}{8}}$, $\lambda_{1,2}(0)=\frac{1}{4}$.}
\end{figure}
\begin{figure}[htba]
{\centering
\resizebox*{0.45\textwidth}{!}{\includegraphics{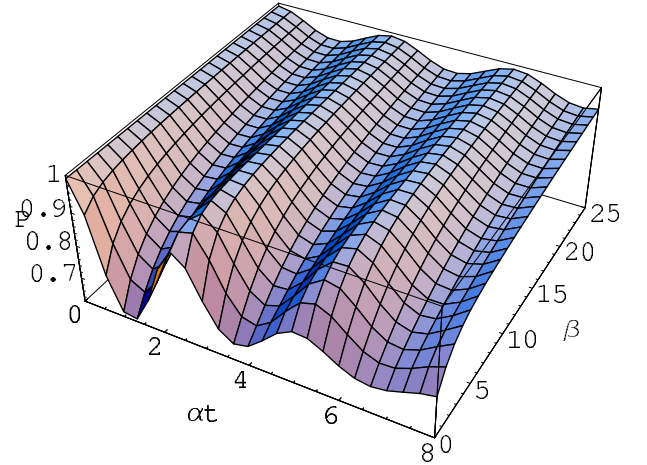}}
\par}
\caption{\label{figure14} (Color online) Evolution in time of the purity for different values of the bath temperature in the case
$N\to\infty$ with
 $\gamma=0$, $g=1$ and $\mu=\alpha$.}
\end{figure}

  Let us now discuss  the short-time behavior of the
reduced dynamics. The aim here is to find simple analytical expressions which describe
the variation of the reduced density matrix at short time scales.
 It is clear from the expressions of the
functions $\eta(t)$, $\zeta(t)$ and $\xi(t)$ that the term of
interest which describes the decay of the Bloch vector components is
given by $\mathrm e^{-\frac{t^2}{(g\beta+2)}}$.
  We shall look for  functions of the form
$\mathrm e^{-\frac{t^2}{(g\beta+2)}}g(t)$ where $g(t)$ is some
complex-valued function of the time . In the case of the
off-diagonal elements,  the ansatz $g(t)=\mathrm e^{2i\mu t}$ can be
justified by the competition of two processes, namely oscillations
due to the external magnetic field and damping due to the coupling
with the environment. In the limiting case where the magnetic field
is absent,  it is found that for small values of the time, the decay
is purely Gaussian. On the other hand if we assume that there is no
coupling between the bath and the central spin, i.e. $\alpha=0$,
then the dynamics is governed by the external magnetic field. The
component $\lambda_3(t)$ is not affected by the magnetic field even
when there is no coupling between the spin and the bath. It is shown
in Ref.~\onlinecite{15} that this component decays two times faster
than the other ones. Consequently, the short-time behavior of the
reduced density matrix can be described by
\begin{align}
\frac{\lambda_{3}(t)}{\lambda_3(0)}&\approx\exp\Bigl(- \frac{2t^2}{2+g\beta}\Bigl)\label{short3},\\
\frac{\lambda_1(t)-i\lambda_2(t)}{\lambda_{1}(0)-i\lambda_2(0)}&\approx\exp\Bigl(-
\frac{t^2}{2+g\beta}+2 i \mu t \Bigl)\label{short4}.
\end{align}
In Fig.~\ref{figure11}, the short time behavior of the Bloch vector components $\lambda_3(t)$
and $\lambda_1(t)$ is shown together with the approximations (\ref{short3}) and (\ref{short4}); these  are in good
agreement with the exact solutions.

\begin{figure}[htba]
{\centering
\resizebox*{0.45\textwidth}{!}{\includegraphics{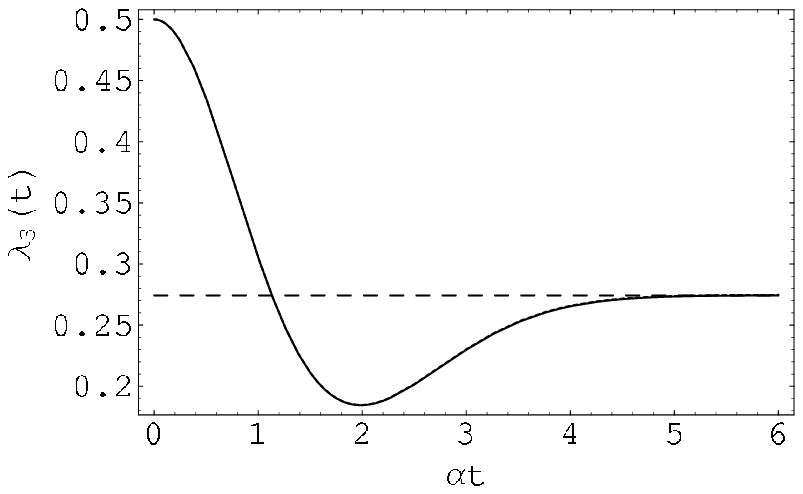}}
\resizebox*{0.45\textwidth}{!}{\includegraphics{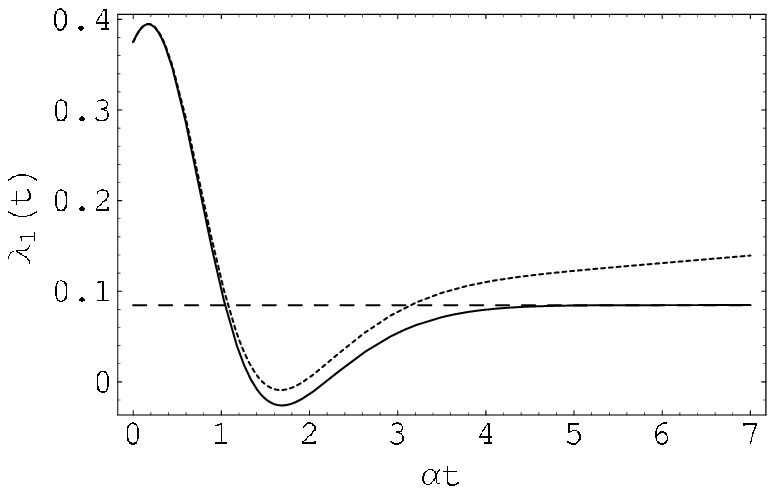}}
\par}

\caption{\label{figure15}  Evolution of $\lambda_3(t)$ and $\lambda_1(t)$  for $N=600$ (dotted lines)
 and $N\to\infty$ (solid lines). The dashed lines correspond to the asymptotic values.
Here, $\gamma=2\alpha$, $g=1$, $\beta=1.5$, $\Delta=1$, $\mu =0.3 \alpha$, $\lambda_3(0)=\frac{1}{2}$ and $\lambda_{1,2}(0)=\frac{3}{8}$. The plots corresponding to $\lambda_3(t)$
are almost identical, the latter component saturates with respect to $N$ faster than $\lambda_1(t)$.}
\end{figure}

There exist many measures that allow for the quantification of the
degree of the decoherence due to the interaction with an
environment. In this work we use the  measure $D(t)=1-P(t)$, where
\begin{equation}
P(t)=\mathrm{tr}\{\rho(t)^2\}
\end{equation}
is the purity of the central system. Note that in the previous expression  the trace is performed
 over
the degrees of freedom of the central spin. The purity takes its
maximum value  $1$ at pure states;
 its minimum value, $1/2$, corresponds to the fully mixed
state $\rho=\mathbf{1}_{2}/2$. In our case, the purity
 can be expressed  in terms of the Bloch vector components
as
\begin{equation}P(t)=\frac{1}{2}\Bigl(1+\lambda_1(t)^2+\lambda_2(t)^2
+\lambda_3(t)^2\Bigl).
\end{equation}
The above expression shows that the  decay of the purity in the
short-time regime described by Eqs.~(\ref{short3})-(\ref{short4}) is
Gaussian which reflects the non-Markovian character of the dynamics.
The decay process is slowed down by decreasing the temperature of
the bath and/or applying a magnetic field of sufficient strength as
illustrated  in Figs.~\ref{figure13} and \ref{figure14}. When
$t\to\infty$, the central spin shows partial decoherence; if
$\mu=0$, then  the asymptotic value of the purity is independent of
the bath temperature as expected (see Fig.~\ref{figure13}).

\subsection{The case  $\gamma\neq 0$} The time dependence of the Bloch vector components when the constant $\gamma$ is
different from zero can be obtained  with the same method used in
the previous subsection. Since $\gamma\neq0$, the quantities
$M_{1,2}$ are $m$-dependent which means that the effect of the
anisotropy constant has to be taken into account.
  When $\gamma\neq0$, we need to perform  double  integration
with respect to the real variables $r$ and $m$ as shown in Eq.~(\ref{lim}).
By making use of  the Riemann-Lebesgue lemma, it is possible
to find  the following
asymptotic expression for the function $\eta(t)$ obtained by replacing $\mu$ by $\mu+\gamma m$ in Eq.~(\ref{fun})(see Fig.~\ref{figure15})
\begin{widetext}
\begin{equation}
\lim\limits_{t\to\infty}\eta(t)=1-\frac{1}{\sqrt{\pi}}(2+g\beta)\sqrt{2+g\beta\Delta}\int\limits_{-\infty}^{\infty}(\mu+\gamma
m)^2 \mathrm e^{(\mu+\gamma m)^2(g\beta+2)-(2+g\beta\Delta)m^2}
\Gamma\Bigl(0,(\mu+\gamma m)^2(2+g\beta)\Bigl) dm. \label{ginf}
\end{equation}
\end{widetext}
\begin{figure}[htba]
{\centering
\resizebox*{0.45\textwidth}{!}{\includegraphics{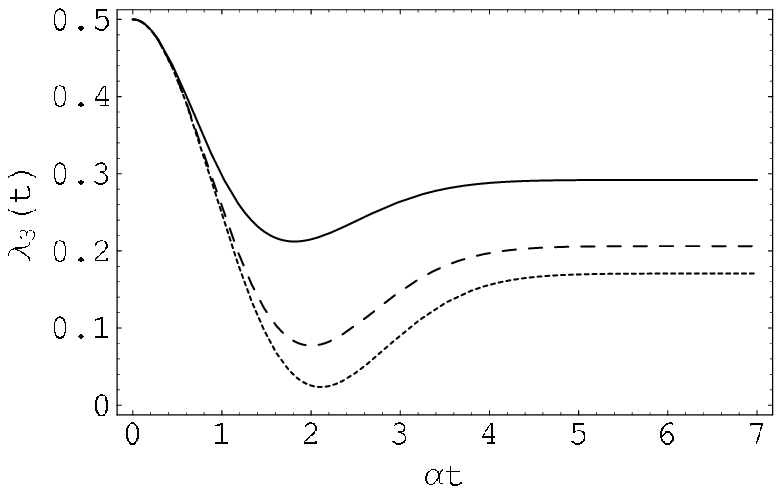}}
\resizebox*{0.45\textwidth}{!}{\includegraphics{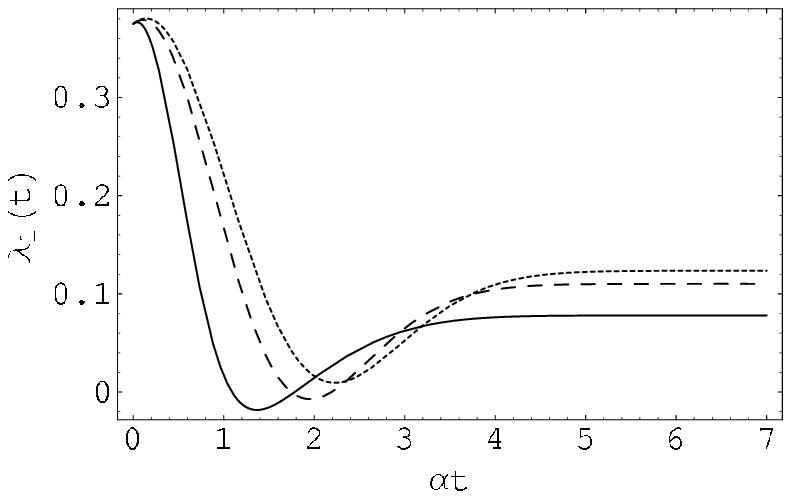}}
\par}

\caption{\label{figure17}  Dependence of $\lambda_3(t)$ and
$\lambda_1(t)$ on the anisotropy constant in the case $N\to\infty$:
$\Delta=0$ (solid lines), $\Delta=5$ (dashed lines),  and
$\Delta=10$ (dotted lines). Here, $\gamma=2\alpha$, $g\beta=1$, $\mu
= 0.1 \alpha$, $\lambda_3(0)=\frac{1}{2}$ and
$\lambda_{1,2}(0)=\frac{3}{8}$.}
\end{figure}

Obviously, the functions $\zeta(t)$ and  $\xi(t)$
tend to zero when $t\to\infty$.  Hence, even if we set $\mu=0$, the asymptotic state is still
temperature dependent. Nevertheless, the dependence of the Bloch vector components on the bath temperature is quiet
 similar to the one corresponding to $\gamma=0$.
 The influence of the magnetic field on the dynamics of the central spin is appreciable only when its strength
 is sufficiently large;
 this can be seen from
the absence of oscillations in the components $\lambda_1(t)$ and $\lambda_3(t)$ displayed in figure \ref{figure15}.
 Fig.~\ref{figure17} shows that the off-diagonal
 elements  decay slower and assume larger asymptotic values when the anisotropy constant $\Delta$ increases.
  The opposite situation holds for the component $\lambda_3(t)$, that is when $\Delta$ decreases
  the latter component assumes larger
  asymptotic limits.
\begin{figure}[htba]
{\centering
\resizebox*{0.45\textwidth}{!}{\includegraphics{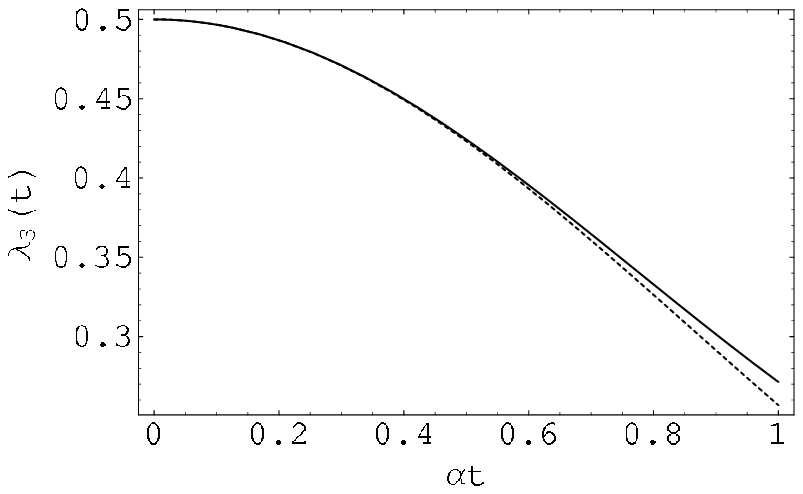}}
\resizebox*{0.45\textwidth}{!}{\includegraphics{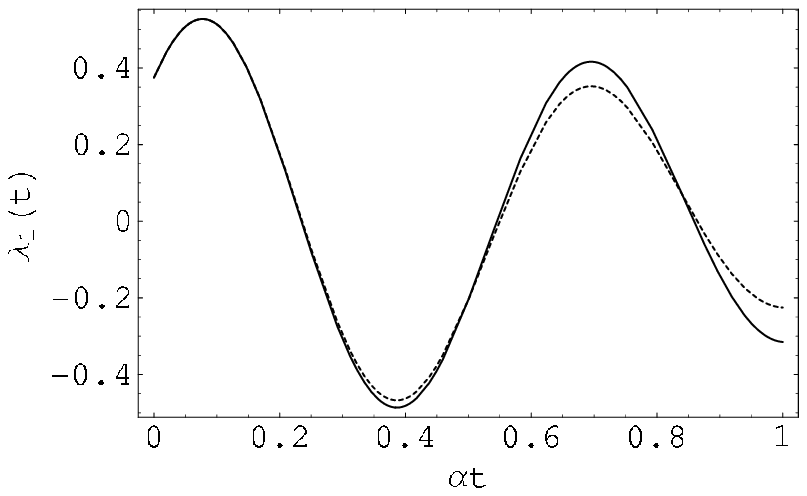}}
\par}

\caption{\label{figure19}  The short-time behavior of $\lambda_3(t)$ and $\lambda_1(t)$ in the case
 $N\to\infty$.
The solid lines correspond to the exact solutions, the the dotted lines denote the  approximations (\ref{short5}-\ref{short6}).
The parameters are  $\gamma=1\alpha$, $\Delta=0$, $g\beta=1$ and $\mu =5\alpha$. The
initial conditions are the same as in Fig.\ref{figure17}.}
\end{figure}

     At short times the diagonal elements of the reduced density matrix do not depend
  on $\Delta$ in contrast with the off-diagonal ones. In the case of the  Heisenberg $XY$ model, i.e. when $\Delta=0$, the short-time behavior of the Bloch
vector components  can
be determined with the same
procedure used in the case where $\gamma= 0$. The main difference here is that the contribution
of the interaction $V=\frac{\gamma}{\sqrt{N}}S^0_zJ_z$  has to be  taken into account. Using the result we obtained in  Eq.~(\ref{gamma}),
we can describe the short-time behavior of the reduced density matrix  by (see Fig.~\ref{figure19})
\begin{align}
\frac{\lambda_{3}(t)}{\lambda_3(0)}&\approx \exp\Bigl(- \frac{2t^2}{2+g\beta}\Bigl)\label{short5},\\
\frac{\lambda_1(t)-i\lambda_2(t)}{\lambda_{1}(0)-i\lambda_2(0)}&\approx \exp\Bigl(-
\frac{t^2}{2+g\beta}-\frac{\gamma^2 t^2}{2}+2 i \mu t\Bigl)\label{short6}.
\end{align}
For $\Delta\neq0$, the situation is much more complicated; here we
only discuss the  special case where
  $\mu=\alpha=0$, and $\Delta>>1$. The last condition implies that the transverse term of $H_B$ can
  be neglected compared to the longitudinal one. Under the above assumption, $H_B$
  simplifies to $g\Delta/N S_z^2$ and thus
   all interactions are of Ising type. Therefore, the operators $H_{SB}$ and $H_B$ commute with each other which means
    that
     the diagonal elements are not affected by the coupling to the environment.
     The coherence of the central spin can be calculated as usual.
  Taking the limit of an infinite number of spins and using the probability density function corresponding to the
  random variable $m$, we find that
  the off-diagonal elements decay according to the Gaussian law $\exp\Bigl[-\frac{\gamma^2 t^2}{2+g\beta\Delta}\Bigl]$.
  Hence the larger the anisotropy constant
  the slower the decay of the off-diagonal elements, which explains the behavior at short times of $\lambda_1(t)$ displayed in
  Fig.~\ref{figure17}. More details about the case of Ising couplings
  can be found in Ref.~\onlinecite{30}.
  To end our discussion about the short-time behavior, it should
   be noted that the deviation of the short-time expressions  (\ref{short4}) and (\ref{short6})
    from the exact solutions
    depends on the value of the
  strength of the magnetic field. For small values of $\mu$, the above relations are valid at relatively
  large intervals of time. However, as $\mu$ increases,
  the domains of time for which the above approximations are valid become shorter.
\begin{figure}[htba]
{\centering
\resizebox*{0.45\textwidth}{!}{\includegraphics{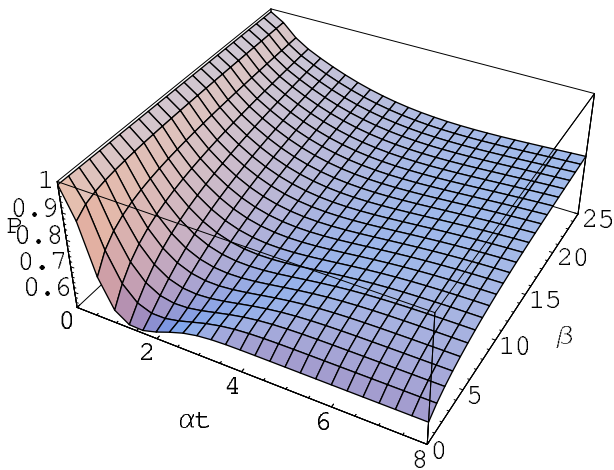}}
\par}

\caption{\label{figure21} (Color online) Purity evolution for different values of the bath temperature
in the case $N\to\infty$ with
 $\gamma=2\alpha$, $\Delta=1$, $g=1$, and $\mu=\alpha$. The
initial conditions are $\lambda_3(0)=\sqrt{\frac{7}{8}}$, $\lambda_{1,2}(0)=\frac{1}{4}$.}
\end{figure}

The variation in time of the purity in this case  differs from the one corresponding to $\gamma=0$ by the
  suppression of the damped oscillations  caused by the external magnetic field as shown in Fig~\ref{figure21}.
  This is mainly due  to  the interaction described by the Hamiltonian $V$. Consequently, the central spin
   decoheres less
  when $\gamma$ is equal to zero. The above result  was expected because the longitudinal coupling vanishes:
    the central spin is less correlated to
   the environment and  thus   the destructive effect of the environment on the coherence of the two-level system
    is less appreciable. Indeed, the decoherence
   time
  constant is  found to be inversely proportional to $\gamma$, namely
  $\tau=\frac{1}{\alpha}\sqrt{\frac{4+2g\beta}{2+\gamma^2(2+g\beta)}}$. This  simply  implies that
   $\tau\to 0$ as $\gamma\to\infty$.

\section{Conclusion}
 In conclusion we have investigated the dynamics of a spin-$\frac{1}{2} $
  particle,  subjected to the effect of  a locally applied external
 magnetic field, and coupled to
  anisotropic Heisenberg  spin environment in thermal equilibrium.
   The reduced density matrix was analytically
 derived for finite number of spins in the environment and arbitrary values of the interaction strengths.
 The evolution in time of the central spin depends on the nature of  interactions within the bath. In the case of
 ferromagnetic environment, the decay of the Bloch vector components is Gaussian accompanied by fast damped oscillations.
  In the antiferromagnetic case, the components of the bloch
vector saturate with
  respect to the number of environmental spins and display partial decoherence. We showed that
the partial trace over the degrees of freedom of the bath
   can be calculated using the convergence of
  the  rescaled bath operators to
   normal  independent Gaussian random variables. This
  allowed us to study the case  of  an infinite number of environmental spins, and to analytically derive
   the asymptotic behavior
  of the components of the Bloch vector.
  The above limit  represents a good approximation for the
 cases with finite number of spins ($N\sim 100$).
 At  short time scales,  the decay of
   the off-diagonal elements
  is found to be Gaussian with a decoherence time  constant given by $\tau=\sqrt{\frac{2+g\beta}{ \alpha^2}}$ ( $\gamma=0$ ).
   This  result is mainly due to
  the non-Markovian nature of the dynamics, which in turn follows from the time-independence
  of the bath correlation functions and the symmetry of the bath
  Hamiltonian. Also, it has be shown  that the  effect of low bath temperatures on the
  decoherence of the central spin is similar to that of strongly applied  magnetic fields and large bath anisotropy
  . The  results obtained
  in this work are valid for
  any number of spins in the environment and arbitrary values of
  the strength of the external magnetic field and the bath temperature.
   They are in good agreement with those of Ref.~\onlinecite{21} where the authors
   studied decoherence of electron spins in
 quantum dots. The model can be generalized to the case of two or more  interacting qubits
  where questions related to the decoherence and the entanglement can be investigated.
\acknowledgments Y. H. would like to express his gratitude for  the
warm hospitality extended to him  during his visit to Institut de
Physique, Universit\'{e} Mentouri-Constantine,  Algeria, where this
work was partially carried out. The financial support from the South
African National Research Foundation within the Focus Area Programme
Unlocking the Future is gratefully acknowledged.

\appendix*
\section{Derivation of the analytical form of $\eta(t)$}

 This appendix is devoted to the  derivation of the asymptotic behavior and the analytical form of the
 function $\eta(t)$ appearing in Eq.~(\ref{eta}). Explicitly we have ($\Delta=0$)
\begin{equation}
\eta(t)=\frac{8}{\bar{Z}} \int\limits_0^\infty \mathrm e^{-(g\beta+2) r^2} \frac{r^2}{\mu^2+r^2}
 \sin^2\Bigl(t \sqrt{\mu^2+r^2} \Bigl) dr.
\end{equation}
By making the following  change of variable $r^2=s^2-\mu^2$ and
taking into account the trigonometric  equality $\sin^2
x=\frac{1}{2}\Bigl[1-\cos(2x)\Bigl]$, we can rewrite the above
function as
\begin{align}
\eta(t)&=\frac{2}{\bar{Z}}\ \mathrm e^{(2+g\beta)\mu^2}\Bigl\{\int\limits_{\mu^2}^\infty  dv^2 \mathrm e^{-(2+g\beta) v^2}\Bigl[1 -\cos(2 v t)\Bigl]
\nonumber \\&-2\mu^2\Bigl[\int\limits_\mu^\infty \frac{dv}{v}
  \mathrm e^{-(2+g\beta)v^2}\Bigl(1-\cos(2 v t)\Bigl)\Bigl]\Bigl\}.
\end{align}
 The Riemann-Lebesgue lemma implies that the  second and the fourth terms involving the cosine function in the
 above expression  vanish when $t\to\infty$.
 The first term can be easily evaluated and we simply get
 \begin{equation}
 \int\limits_{\mu^2}^\infty  dv^2 \mathrm e^{-(2+g\beta) v^2}= \frac{1}{2+g\beta} \mathrm e^{-(2+g\beta)\mu^2}.
\end{equation}
  The third term reads
 \begin{equation}
2\int\limits_\mu^\infty \frac{dv}{v} \mathrm e^{-(2+g\beta) v^2}=\int\limits_{\mu^2(2+g\beta)}^\infty \frac{dv^2}{v^2} \mathrm e^{- v^2}=\Gamma\Bigl(0,(2+g\beta)\mu^2\Bigl),
\end{equation}
where we have made the change of variable $(2+g\beta)v^2\to v^2 $. Taking into account the expression of $\bar{Z}$
in Eq.~(\ref{z}) we obtain the asymptotic expression of $\eta(t)$ displayed in Eq.~(\ref{assym}).

 The second term simplifies to
 \begin{widetext}
 \begin{align}
 \mathrm {Re}\ \Bigl\{2\ \mathrm e^{-\frac{ t^2}{2+g\beta}}\int\limits_\mu^\infty v \ d v \ \mathrm \exp\Bigl[-(2+g\beta)\Bigl(v+\frac{i t}{2+g\beta}\Bigl)^2 \Bigl]
 \Bigl\}=e^{-\frac{t^2}{2+g\beta}}\mathrm{Re}\Bigl\{ \underbrace{\int\limits_{\delta^2}^\infty dv \frac{ \mathrm e^{-v}}{2+g\beta}}\limits_{I_1}
 -\underbrace{\frac{ 2 i t}{(2+g\beta)^{\frac{3}{2}}}
 \int\limits_{\delta}^\infty dv \ e^{-v^2}}\limits_{I_2}\Bigl\},
 \end{align}
 \end{widetext}
 where $\delta=\sqrt{2+g\beta}(\mu+\frac{i t}{2+g\beta})$. One can easily check that
\begin{equation}
 \mathrm {Re}\ (I_1)=\frac{1}{2+g\beta} \exp\Bigl[-(2+g\beta) \mu^2+\frac{t^2}{2+g\beta}\Bigl] \cos\Bigl(2 \mu t\Bigl).
 \end{equation}
The second integral is given by the complementary error function, namely
\begin{equation}
 I_2=\frac{\sqrt{\pi} i t}{(2+g\beta)^{\frac{3}{2}}}    \mathrm {erfc}\Bigl[\Bigl(\mu+\frac{i t}{2+g\beta}\Bigl)
\sqrt{2+g\beta}\Bigl].
\end{equation}
It is then sufficient to use the property $2\ \mathrm{Im} \ \mathrm {erfc}(a+i t)= i\Bigl[\mathrm {erf}(a +i t)-\mathrm {erf}(a-it)\Bigl]$,
where $a$ is real and $\mathrm{Im}(x)$ stands for the imaginary part of $x$,
to
get the first three terms appearing in the right-hand side of Eq.~(\ref{eta}).

Similarly, we have
\begin{align}
\mathrm{Re}\Bigl\{ \ 2\int\limits_{\mu}^\infty \frac{dv}{v} \exp\Bigl[-(2+g\beta)\Bigl(v+\frac{it}{2+g\beta}\Bigl)^2\Bigl]\Bigl\}\nonumber \\
=\mathrm{Re} \Bigl\{\ 2\int\limits_{\delta}^\infty\frac{ds}{s-\frac{it}{\sqrt{2+g\beta}}} \mathrm e^{-s^2}\Bigl\},
\end{align}
where we have introduced the new variable $s=(v+\frac{it}{\sqrt{2+g\beta}})\sqrt{2+g\beta}$. By multiplying the numerator and the denominator
of the quantity under the sign of integral  by $s+\frac{it}{\sqrt{2+g\beta}}$ we get two new integrals.
The first one is given by
\begin{align}
\mathrm{Re}\Bigl\{ 2\int\limits_{\delta}^\infty s\ ds \frac{\mathrm e^{-s^2}}{s^2+\frac{t^2}{2+g\beta}}\Bigl\}&= \mathrm
e^{\frac{t^2}{2+g\beta}}\mathrm{Re}\Bigl\{ \int
\limits_{\delta_2}^\infty  \frac{ds}{s} \mathrm e^{-s}\Bigl\}\nonumber \\ &=\exp\Bigl\{\frac{t^2}{2+g\beta}\Bigl\}\mathrm{Re}\Bigl
\{ \Gamma(0,\delta_2)\Bigl\},
\end{align}
where $\delta_2=(2+g\beta)\mu^2+2 \mu i t$. The remaining integral  defines the function $\mathcal M$, namely
\begin{align}
\mathcal {M}(t;\mu,\beta)&=\exp\Bigl\{-\Bigl[\frac{t^2}{2+g\beta}-(2+g\beta)\mu^2\Bigl]\Bigl\}\nonumber\\
&\times \mathrm{Re} \Bigl\{ 2it \sqrt{2+g\beta}\int\limits_{\delta}^
\infty\frac{\mathrm e^{-s^2}}{
s^2+\frac{t^2}{2+g\beta}}\ ds\Bigl\}.\label{M}
\end{align}
The analytical expressions of the functions $\xi(t)$ and $\zeta(t)$ can be determined with the same method.
In the case $\gamma\ne0$ we should replace $\mu$ by $\mu+\gamma m$ and then perform the integration with respect to $m$.
For practical investigation, numerical integration is used.

\end{document}